\documentclass[11pt,a4paper]{article}
\usepackage{jheppub}

\usepackage{bbm,amsmath,graphicx,amssymb,mathrsfs}
\usepackage{mathtools}
\usepackage{epstopdf}
\usepackage{bm}
\usepackage{verbatim}
\usepackage{slashed}
\usepackage{multirow}

\usepackage{graphicx}
\usepackage{empheq}

\graphicspath{{./}}

\usepackage{subcaption}

\newcommand{\bea}{\begin{eqnarray}}
\newcommand{\eea}{\end{eqnarray}}
\newcommand{\bean}{\begin{eqnarray*}}
\newcommand{\eean}{\end{eqnarray*}}
\newcommand{\nn}{\nonumber \\}

\def\Label#1{\label{#1}%
  \smash{\hbox to0pt{\raise1ex\hbox{\tiny[#1]}\hss}}}

\newcommand{\mc}{\big|_{\rm m.c.}}

\def\nn{\nonumber}

\def\stamp{--- {\bf \today} --- {\bf \jobname.tex}}

\def\fs_#1{\mathfrak{s}(#1)}

\def\BE{\begin{equation}}
\def\EE{\end{equation}}
\def\spa#1.#2{\left\langle#1\,#2\right\rangle}
\def\spb#1.#2{\left[#1\,#2\right]}
\def\lor#1.#2{\left(#1\,#2\right)}

\newcommand\fverb{\setbox\fverbbox=\hbox\bgroup\verb}
\newcommand\fverbdo{\egroup\medskip\noindent%
            \fbox{\unhbox\fverbbox}\ }
\newcommand\fverbit{\egroup\item[\fbox{\unhbox\fverbbox}]}
\newbox\fverbbox

\title{Maximal Cuts in Arbitrary Dimension}

\author[a]{Jorrit Bosma,}
\author[b,c]{Mads Sogaard,}
\author[a]{Yang Zhang}

\affiliation[a]{Institute for Theoretical Physics, ETH Z\"urich, CH 8093 Z\"urich, Switzerland}
\affiliation[b]{PRISMA Cluster of Excellence, Johannes Gutenberg University, D 55099 Mainz, Germany}
\affiliation[c]{SLAC National Accelerator Laboratory, Stanford University, Menlo Park, CA 94025, USA}

\emailAdd{jbosma@itp.phys.ethz.ch, sogaard@slac.stanford.edu, yang.zhang@phys.ethz.ch}
\abstract{We develop a systematic procedure for computing maximal unitarity cuts
of multiloop Feynman integrals in arbitrary dimension. Our approach is based on
the Baikov representation in which the structure of the cuts is particularly
simple. We examine several planar and nonplanar integral topologies and
demonstrate that the maximal cut inherits IBPs and dimension shift identities
satisfied by the uncut integral. Furthermore, for the examples we calculated, we
find that the maximal cut functions from different allowed regions, form
the Wronskian matrix of the differential equations on the maximal cut.}


\begin{document}
%%%%%%%%%%%%%%%%%%%%%%%%%%
\maketitle

%%%%%%%%%%%%%%%%%%%%%
\section{Introduction}
%%%%%%%%%%%%%%%%%%%%%%%
Quantum field theory scattering amplitudes are mathematical quantities enabling
physicists to make predictions for physical observables in high energy particle
experiments such as the Large Hadron Collider (LHC) at CERN. Although scattering
amplitudes are among the most important objects in this research direction,
sufficient precision frequently requires explicit computations that are extremely
challenging, even with powerful modern techniques. The reason is often the
complexity of the Feynman integrals involved, and an inadequate understanding of
the underlying mathematics and the surprisingly rich hidden structures of
scattering amplitudes that are continuously being unravelled.

The past two decades have seen enormous progress in the development of new
enhanced methods for computing multiloop scattering amplitudes. The traditional
techniques due to Feynman are no longer preferred by experts for
state-of-the-art calculations. Instead, scattering amplitudes are typically
reduced to a linear combination of integrand or integral basis elements, whose
coefficients then become the primary quantities of interest after the integrals
have been carried out once and for all. All one-loop integrals can be expressed
in terms of simple algebraic functions along with the logarithm and dilogarithm,
whose arguments are again algebraic functions. Today, fully automated
computation of one-loop amplitudes has been achieved, either via the unitarity
method \cite{Bern:1994zx,Bern:1994cg} and its refinements
\cite{Britto:2004nc,Forde:2007mi}, or by the Ossola, Papadopoulos and Pittau
(OPP) approach \cite{Ossola:2006us,Ossola:2008xq} at the level of the integrand.
More recently, extensions of these techniques to two loops in general theories
have been reported, forming the frontier of next-to-next-to-leading (NNLO)
corrections. A key element in these developments has been the application of
(computational) algebraic geometry \cite{Zhang:2016kfo}. See
refs.~\cite{Mastrolia:2011pr,Badger:2012dp,Mastrolia:2012wf,Badger:2013gxa,
Badger:2015lda,Badger:2016ozq,Mastrolia:2016dhn,Badger:2017gta} for the multiloop version of the OPP method, and
refs.~\cite{Kosower:2011ty,Johansson:2012zv,Johansson:2013sda,
Sogaard:2013yga,Sogaard:2013fpa,Sogaard:2014ila,Sogaard:2014oka,
Sogaard:2014jla,Johansson:2015ava,Abreu:2017idw,Abreu:2017xsl} for progress on direct extraction of integral
coefficients from an integral basis.

One of the remaining bottlenecks in the unitarity and OPP based methods at the
multiloop level is the computation of the Feynmans integrals themselves.
Morover, at two loops and beyond it is more complicated to even determine an
appropriate integral basis \cite{Gluza:2010ws} at higher multiplicity. Given a
complete set of master integrals for the problem in consideration, the standard
procedure for evaluating them is to derive differential equations
\cite{Kotikov:1991pm,Kotikov:1990kg,Bern:1992em,Remiddi:1997ny,Gehrmann:1999as,
Ablinger:2015tua}
in the external kinematic invariants, reduce the resulting expression using
integration-by-parts (IBP) identities \cite{Chetyrkin:1981qh} to form a linear
system of equations. The integrated expressions are constructed from a much less
restricted class of transcendental functions, including for instance generalized
polylogarithms. These ideas have proven extremely useful in practice over the
years. In particular, if the basis integrals are chosen properly, the
differential equations are brought to the canonical form
($\epsilon$-form) proposed by Henn
\cite{Henn:2013pwa}, leading to significant simplifcations. (See
ref.~\cite{Henn:2014qga} for a pedagogical review and 
refs.~\cite{Argeri:2014qva,Lee:2014ioa,Meyer:2016slj,Prausa:2017ltv,Gituliar:2017vzm,Adams:2017tga}
for algorithms and packages for finding the canonical form.)

Motivated by the tremendous success of generalized unitarity at one loop, we
present here a systematic strategy for evaluating maximal cuts of multiloop
Feynman integrals, properly defined within dimensional
regularization. $D=4$ two-loop maximal unitarity was first achieved by
the elegant contour method in ref.~\cite{Kosower:2011ty} by Kosower
and Larsen, and then generalized to other $4D$ integrals with
external or internal massive legs, nonplanar topology and three-loop
order \cite{Johansson:2012zv,Johansson:2013sda,
Sogaard:2013yga,Sogaard:2013fpa,Sogaard:2014ila,Sogaard:2014oka,
Sogaard:2014jla,Johansson:2015ava}. The work by
Frellesvig and Papadopoulos \cite{Frellesvig:2017aai} studied the
$D$-dimensional maximal cut via the Baikov representation \cite{Baikov:1996cd,Baikov:1996rk,Baikov:2005nv}, and
also explicitly presented the $\epsilon$ expansion around $4D$ for the maximal cut
function. On the other hand, using the multivariate residue calculus of Leray, refs.~\cite{Abreu:2017ptx, Abreu:2017enx} give a precise
definition of the cut Feynman integral in dimensional
regularization, and argue that integral relations carry over from the uncut Feynman integrals to the cut integrals. 
While refs.~\cite{Abreu:2017ptx, Abreu:2017enx} focus on the one-loop case, they predict the construction works at higher loops as well.
%This beautiful construction is predicted to work at higher
%loops as well \cite{Abreu:2017ptx, Abreu:2017enx}.  

In this paper, we systematically study maximal cuts in any spacetime
dimensions, by computing the Baikov integrals on the maximal cut over all
possible regions, and verify that the maximal cut functions
automatically incorporate all integral relations such as IBPs, dimension
recurrence relations, and differential equations on the maximal cut. This method applies equally well to planar and nonplanar integrals
with and without massive particles. A careful treatment lends credence to the
belief that for an integral topology with $m$ master integrals,  {\it
  each} master integral would have precisely
$m$ linearly independent maximal cut functions in $D$ dimensions. We provide
nontrivial evidence that these $m$ maximal cut functions form the Wronskian
matrix associated with the differential equation satisfied by the master
integrals on the maximal cut. The leading terms of this Wronskian matrix are
useful to transform the differential equation to the canonical form.

This paper is related to ref.~\cite{Primo:2016ebd} by Primo and
Tancredi, and ref.~\cite{Zeng:2017ipr} by Zeng, which study the
differential equations on the maximal cut systematically,
and ref.~\cite{Frellesvig:2017aai} by Frellesvig and Papadopoulos
which applies the efficient loop-induction
method to find the maximal cut in the Baikov representation. We remark
that our paper is characterized by (i) always retaining complete dimension dependence for all cuts in
closed form, so the limit behaviour of the cut function near any
integer dimension can be easily obtained, (ii) giving full dependence of the irreducible scalar product (ISP)
indices to make all integral relations (like IBP relations) manifest,
(iii) most importantly, providing the {\it complete solution system} (Wronskian matrix) for
$D$-dimensional IBPs, dimension recurrence identities and differential
equations on the maximal cut, from an analysis of all allowed
integration regions of the Baikov representation on the cut. 

This paper is organized as follows: in section 2, we present the
Baikov integral representation with the maximal cut and integral regions
for real kinematics. In section 3 and 4, we review the simple
$D$-dimensional maximal cut examples with zero or one ISP. Section 5
and 6 contain our main examples, for which the Baikov integration on
the cut over different regions gives independent solutions for on-shell
IBPs, dimension recurrence relations and differential equations. We
explicitly show these solutions are complete by studying the Wronskian
of the differential equation. 

%%%%%%%%%%%%%%%%%%%%%
\section{Baikov representation and maximal cuts}
We are interested in $L$-loop Feynman integrals with $n$ external
momenta and $k$ propagators,
\begin{equation}
  \label{FI}
  I_{a_1,\ldots a_k, a_{k+1}, \ldots a_n}\equiv \int
  \frac{d^D l_1}{\pi^{D/2}} \ldots \frac{d^D l_L}{\pi^{D/2}}
  \frac{1}{D_1^{a_1}\ldots D_k^{a_k} D_{k+1}^{a_{k+1}} \ldots D_m^{a_{m}} }\,,
\end{equation}
where $m=(n-1)L+L(L+1)/2$ and $k\leq m$. $D_1$, $\ldots ,$ $D_k$ are denominators of Feynman
propagators, and $D_{k+1}$, $\ldots ,$ $D_{m}$ are the irreducible
scalar products (ISPs). So we require that (for integrals in this
particular sector),
\begin{align}
  \label{eq:33}
&a_i>0\,,\quad  i=1,\ldots k\,,\nonumber\\
&a_j\leq 0\,,\quad j=k+1,\ldots m \,.
\end{align}

We use the Baikov representation \cite{Baikov:1996rk,Baikov:1996cd,Baikov:2005nv} of
\eqref{FI}. Schemetically, 
\begin{equation}
  \label{eq:34}
  I_{a_1,\ldots a_k, a_{k+1}, \ldots a_n}=C(D,x) \int_A  dz_1
  \ldots dz_m \ \frac{ F(z)^{\frac{D-L-n}{2}}}{z_1^{a_1}\ldots z_k^{a_k} z_{k+1}^{a_{k+1}} \ldots z_m^{a_{m}} }\,,
\end{equation}
where $F(z)$ is the Baikov polynomial. The overall factor $C(D,x)$ is a
product of hypersphere areas, the Jacobian of the Baikov transformation
and the Gram determinant. The kinematic variables are collectively
called $x$. 

In this paper, we simply consider real-valued external and internal
momenta to simplify the discussion of the Baikov integration region
$A$. For real momenta, $A$ is determined by the spacetime
metric signature and Cauchy-Schwarz inequality. For $L=1$, the integration
region $A$ is simply defined by $F(z)\geq 0$. For $L=2$, the
integration region $A$ is defined by $F(z)\geq 0$, $\mu_{11}(z)\geq 0$ and
$\mu_{22}(z)\geq 0$, where $\mu_{11}(z)$ and $\mu_{22}(z)$ are defined
as following: the loop momenta are separated into the projections in
the $(n-1)$-dimensional space (spanned by external momenta) and the
orthogonal complement,
\begin{equation}
  \label{eq:35}
  l_1 =\bar l_1 + l_1^\perp,\quad l_2 =\bar l_2 + l_2^\perp\,.
\end{equation}
The inner products of $l_1^\perp$ and $l_2^\perp$ are $\mu_{11}\equiv
-(l_1^\perp)^2$, $\mu_{22}\equiv -(l_2^\perp)^2$, $\mu_{12}\equiv - (l_1^\perp)\cdot (l_2^\perp)$. In terms of the
Baikov representation, the $\mu$'s become polynomials in $z$'s. For real
internal momenta, $\mu_{11}\geq 0$, $\mu_{22}\geq 0$. Furthermore, by
the  Cauchy-Schwarz inequality, $F(z)= \mu_{11}(z)
\mu_{22}(z)-\mu_{12}(z)^2\geq 0$. 

Unitarity cuts become manifest in the Baikov representation. For
example, the maximal cut in Baikov representation is to consider the multivariate residue
at $z_1=z_2=\ldots=z_k=0$ \cite{Baikov:1996rk,Baikov:1996cd, Baikov:2005nv,Lee:2014tja, Ita:2015tya,Larsen:2015ped, Frellesvig:2017aai}. If the integral has no double
propagators, i.e., $a_1=a_2=\ldots=a_k=1$, the maximal cut becomes
\begin{equation}
  \label{mc1}
  C(D,x)\int_{\Omega}  dz_{k+1}
  \ldots dz_m \ \frac{ F(z_{k+1},\ldots z_{m})^{\frac{D-L-n}{2}}}{ z_{k+1}^{a_{k+1}} \ldots z_m^{a_{m}} }\,,
\end{equation}
where $\Omega$ is the intersection of $A$ and the hyperplane
$z_1=z_2=\ldots=z_k=0$. For the case with some $a_i>1$, $1\leq i\leq
k$, derivates of the Baikov polynomial are
needed to get the residue. This form can be used to derive 
integration-by-parts (IBP) identities 
\cite{Ita:2015tya, Larsen:2015ped} and differential equations
\cite{Zeng:2017ipr} on the maximal cut, and to identify master integrals
\cite{Lee:2013hzt, Georgoudis:2016wff}, by using Morse theory, tangent vectors and
syzygy computations. More generally, the non-maximal cut in Baikov
form can be used to derive the complete set of IBPs \cite{Ita:2015tya, Larsen:2015ped}.
 
 In this paper, we systematically study \eqref{mc1} in detail. We find
that frequently the region $\Omega$ decomposes into several
subregions, (see figure \ref{dbox_region} for the subregions of
massless double box as an explicit example)
\begin{equation}
  \label{eq:37}
  \Omega= \Omega_1 \cup \ldots \cup \Omega_s\,,
\end{equation}
where on the boundary $\partial \Omega_j$ of each subregion $\Omega_j$, $F=0$. 
We denote by $s$ the number of such subregions. Then
we can explicitly carry
out the integration on each $\Omega_j$ and apply analytic
continuation in $x$ and $D$. The resulting function is named as the maximal cut function
on the subregion $\Omega_j$,
\begin{equation}
  \label{mc}
  I_{1,\ldots 1, a_{k+1}, \ldots a_n} \mc^{(j)} \equiv C(D,x)\int_{\Omega_j}  dz_{k+1}
  \ldots dz_m \ \frac{ F(z_{k+1},\ldots z_{m})^{\frac{D-L-n}{2}}}{ z_{k+1}^{a_{k+1}} \ldots z_m^{a_{m}} }\,.
\end{equation}

Since $F=0$ on $\partial \Omega_j$, the possible surface term from the
integration of a total derivative vanishes. Hence it is clear that
for each fixed $j$, the functions $I_{1,\ldots 1, a_{k+1}, \ldots a_n}
\mc^{(j)}$ satisfy (the same form of)
integration-by-parts identities on the maximal cut. Similarly, for each fixed $j$, $I_{1,\ldots 1, a_{k+1}, \ldots a_n}
\mc^{(j)}$'s satisfy (the same form of) dimension shift identities and
differential equations on the maximal cut. 

The integrals over these subregions may not be independent.
For each $j$, we may consider $I_{1,\ldots 1, a_{k+1}, \ldots a_n}
\mc^{(j)}$ as a vector with an infinite number of components, indexed by
non-positive integer tuples
$(a_{k+1},\ldots, a_m)$. Let $\mathfrak d$ be the dimension of the
vector space spanned by these $s$ vectors (with
meromorphic functions in $D$ as coefficients).  We then define the
maximal cut as the linear basis of these $s$ vectors,
\begin{empheq}[box=\fbox]{align}
I_{1,\ldots 1, a_{k+1}, \ldots a_n}
\mc\equiv(I_{1,\ldots 1, a_{k+1}, \ldots a_n}
\mc^{(b_1)},\ldots, I_{1,\ldots 1, a_{k+1}, \ldots a_n}
\mc^{(b_{\mathfrak d})})
\,,
\end{empheq}
where $b_1,\ldots, b_{\mathfrak d}$ are the indices of the vectors in
the basis. This is our main formula of this paper.

For the examples we considered in this paper, we find that $\mathfrak d=n_\text{MI}$, the number of
master integrals on the maximal cut. Furthermore, let $S$ be the $\mathfrak d \times
\mathfrak d$ matrix, whose element in the $i$th-row and $j$th-column
is the $i$-th master integral evaluated on the subregion
$\Omega_{b_j}$ of the maximal cut. We find explicitly that, for
the examples we considered, $S$ is the
Wronskian matrix for the differential equation on the maximal cut.  

It is also interesting to study the expansion of $S$ near $D=4$ (or
any integer spacetime dimension). Define $D=4-2\epsilon$. For
example, the expansion is directly related to the $\epsilon$-form of
the differential equation \cite{Henn:2013pwa, Henn:2014qga} on the
maximal cut level
\cite{ Primo:2016ebd, Frellesvig:2017aai,Zeng:2017ipr}.
Suppose that for the $i$-th column of $S$, $S_i$, this expansion reads
\begin{equation}
  \label{eq:32}
  S_i= T_i\cdot  (D-4)^{h_i} + o\big( (D-4)^{h_i}\big)\,,
\end{equation}
where $T_i$ is the leading coefficient column vector, which is itself
$D$ independent. Let $T=(T_1,\ldots, T_{\mathfrak d})$ be the square
matrix consisting of the leading coefficients. If the
differential equation on the maximal cut reads,
\begin{equation}
  \label{eq:36}
  \frac{\partial }{\partial x} I = (A + B \epsilon) I\,,
\end{equation}
where $A$ and $B$ are $\epsilon$ independent, then 
\begin{equation}
  \label{eq:38}
  \frac{\partial }{\partial x} T = A T\,
\end{equation}
by the $\epsilon$-expansion of the differential equation. If $T$ is
invertible\footnote{If $T$ is not invertible, then we may study the null vectors of $T$
and the next-leading expansion coefficients of $S$ to get the transformation matrix.} then the new basis $\tilde I=T^{-1} I$ satisfies
the $\epsilon$-form of the differential equation on the maximal cut,
\begin{equation}
  \label{eq:39}
   \frac{\partial }{\partial x} \tilde I = \epsilon (T^{-1} BT)\tilde I\,.
\end{equation}
This is equivalent to the Magnus rotation in \cite{Argeri:2014qva}.

%%%%%%%%%%%%%%%%%%%%%
\section{Maximal cuts without ISP}
%%%%%%%%%%%%%%%%%%%%%%%
Our first example is the one-loop box integral in $D$ dimensions with purely
massless kinematics. Let $k_1,\dots,k_4$ be the external momenta subject to the
conditions $k_i^2 = 0$ and $\sum_{i=1}^4 k_i = 0$. We define the two independent
Mandelstam invariants by $s = (k_1+k_2)^2$ and $t = (k_1+k_4)^2$. In order to
simplify the problem we will study the maximal cut of the integral rather than
the full integrated expression. There are no ISPs in this case, so we will
instead consider integrals with arbitrary nonnegative powers $a_1,\dots,a_4$ of
the four propagators,
\begin{align}
I_{a_1,a_2,a_3,a_4} = 
\int\frac{d^Dl_1}{\pi^{D/2}}
\frac{1}{D_1^{a_1}D_2^{a_2}D_3^{a_3}D_4^{a_4}}\,.
\end{align}
The denominator factors are given by
\begin{align}
D_1 = l_1^2\,, \quad D_2 = (l_1-k_1)^2\,, \quad 
D_3 = (l_1-k_1-k_2)^2\,, \quad D_4 = (l_1+k_4)^2\,.
\end{align}
A constructive way of proceeding is to examine the Baikov representation. As
discussed in the previous section, the $D$-dimensional scalar box can be written
as the four-fold integral
\begin{align}
I_{1,1,1,1} = 
\frac{2}{\pi^{3/2}\Gamma\big(\frac{D-3}{2}\big)\sqrt{\det G_3}}
\int\prod_{i=1}^4\frac{dz_i}{z_i}F(z)^{\frac{D-5}{2}}\,,
\end{align}
where $F(z)$ is the Baikov polynomial. %\eqref{EQ_BAIKOV_BOX}. 
The maximal cut in
arbitrary dimension is the quadruple cut realized by the replacement
$z_i^{-1}\to\delta(z_i)$ for $i = 1,2,3,4$. This cut localizes the box integral
completely. The value of the maximal cut is thus basically determined by the
Baikov kernel evaluated at the origin,
\begin{align}
I_{1,1,1,1}\mc = 
\frac{2^{7-D}}{\pi^{3/2}\Gamma\big(\frac{D-3}{2}\big)}
s^{\frac{D}{2}-3}t^{\frac{D}{2}-3}(s+t)^{2-\frac{D}{2}}\,.
\label{EQ_BAIKOV_1LOOP_MC}
\end{align}
The leading singularity evaluated in strictly integer dimensions has proven
extremely useful when searching for and designing integrals that have uniform
degree of transcendentality. %see e.g.~\cite{} and references
                             %therein. 
Our
compact analytic expression embodies the well-known result for the leading
singularity in strictly four dimension, but also in odd dimensions, for example
for $D = 5$, %\cite{HennSogaard}, 
\begin{align}
I_{1,1,1,1}^{D=4}\mc = \frac{1}{st}\,, \quad
I_{1,1,1,1}^{D=5}\mc = \frac{1}{\sqrt{s}\sqrt{t}\sqrt{s+t}}\,.
\label{EQ_BOX_LEADING_SINGULARITY}
\end{align}

We can gain further insight by taking advantage of the maximal cut
\eqref{EQ_BAIKOV_1LOOP_MC} to, for example, extract information about IBP
relations for integrals with doubled propagators. It is straightforward to
see that
\begin{align}
I_{1,1,1,2}\mc = 
\frac{2^{D-7}}{\pi^{3/2}\Gamma\left(\frac{D-3}{2}\right)}
(D-5)s^{\frac{D}{2}-3}t^{\frac{D}{2}-4}(s+t)^{2-\frac{D}{2}}\;.
\label{EQ_BAIKOV_1LOOP_DOUBLED_MC}
\end{align}
Upon comparison of eqs.~\eqref{EQ_BAIKOV_1LOOP_MC} and
\eqref{EQ_BAIKOV_1LOOP_DOUBLED_MC} the common Gamma function can be dropped and
therefore we immediately deduce the first very simple instance of an IBP
relation,
\begin{align}
I_{1,1,1,2} = {} & -\frac{D-5}{t}I_{1,1,1,1}+\cdots\,,
\end{align}
where integrals with fewer than four propagators are truncated. More generally,
for generic values of the indices $a_1,\dots,a_4\geq 0$,
\begin{align}
I_{a_1,a_2,a_3,a_4}\mc =
\frac{1}{\Gamma\big(\frac{D-3}{2}\big)}\frac{1}{\sqrt{\det G_3}}
\bigg\{\prod_{i=1}^4\frac{1}{(a_i-1)!}
\frac{\partial^{a_i-1}}{\partial z_i^{a_i-1}}
F(z)^{\frac{D-5}{2}}\bigg\}\bigg|_{z\to 0}\;.
\end{align}
Therefore we can with almost no effort derive any desired IBP identity, for
instance for integrals with tripled or several repeated propagators, simply by
taking multiple derivatives and relating the resulting expression to
eq.~\eqref{EQ_BAIKOV_1LOOP_MC},
\begin{align}
I_{3,1,1,1} = {} & +\frac{(D-6)(D-5)}{2s^2}I_{1,1,1,1}+\cdots\,, \\
I_{2,2,1,1} = {} & +\frac{(D-6)(D-5)}{st}I_{1,1,1,1}+\cdots\,, \\
I_{2,1,1,3} = {} & -\frac{(D-7)(D-6)(D-5)}{2st^2}I_{1,1,1,1}+\cdots\;.
\end{align}

It is also worthwhile to investigate the dimensional dependence of the maximal
cut, that is, examine the dimension shifting and dimensional reduction
identities. The dimension shifting identity relates a $D$-dimensional integral
with an extra-dimensional numerator insertion of $\mu^2$ and a scalar integral
in $(D+2)$ dimensions. At the level of maximal cuts we readily observe that
\begin{align}
I_{1,1,1}^D[\mu^{2r}]\mc = 
\frac{\Gamma\big(\frac{D-4}{2}+r\big)}{\Gamma\big(\frac{D-4}{2}\big)}
I_{1,1,1}^{D+2r}[1]\mc\;.
\end{align}
The ratio of the maximal cuts in this equation is merely a polynomial function
of $D$ as it of course should be. In order to appreciate this fact, recall that
the Gamma function satisfies the functional identity,
\begin{align}
\Gamma(1+z) = z\Gamma(z)\;.
\label{EQ_GAMMA}
\end{align}
Indeed, by iteration of eq.~\eqref{EQ_GAMMA} we get the Pochhammer symbol,
commonly referred to as the ascending factorial,
\begin{align}\label{eq:Pochhammer}
(z)_r\equiv\frac{\Gamma[z+r]}{\Gamma[z]} = z(z+1)(z+2)\cdots(z+r-1)\;.
\end{align}
The dimensional reduction identity relating a $(D+2)$-dimensional integral to a
linear combination of $D$-dimensional scalar integrals is again the Gamma
function property \eqref{EQ_GAMMA} in disguise. For example,
\begin{align}
I_{1,1,1,1}^{D+2} = \frac{st}{2(s+t)(D-3)}I_{1,1,1,1}^D+\cdots\;.
\end{align}

Finally we look at differential equations and the maximal cut. Since we have
access to the leading singularities we can make an educated choice for the
normalization which leads to a canonical differential equation in the spirit of
ref.~\cite{Henn:2013pwa}. Specializing to $D = 4-2\epsilon$ and defining 
$J = stI_{1,1,1,1}$ we find the expected $\epsilon$-form,
\begin{align}
\frac{\partial}{\partial x}J\mc = 
-\frac{\epsilon}{x+1}J\mc\,,
\end{align}
where $x = t/s$. The exact same calculation goes through for $D = 5-2\epsilon$,
where according to eq.~\eqref{EQ_BOX_LEADING_SINGULARITY} we should instead
choose the master integral to be $J = \sqrt{s}\sqrt{t}\sqrt{s+t}I_{1,1,1,1}$.

All results obtained in this section are consistent with the literature.

%%%%%%%%%%%%%%%%%%%%%
\section{Maximal cuts with one ISP}
%%%%%%%%%%%%%%%%%%%%%%%

The next simplest $D$-dimensional maximal cut is the integral with
one ISP. In this section, we briefly review this case. 

\begin{figure}[!h]
    \centering
    \begin{subfigure}[b]{0.3\textwidth}
        \includegraphics[width=\textwidth]{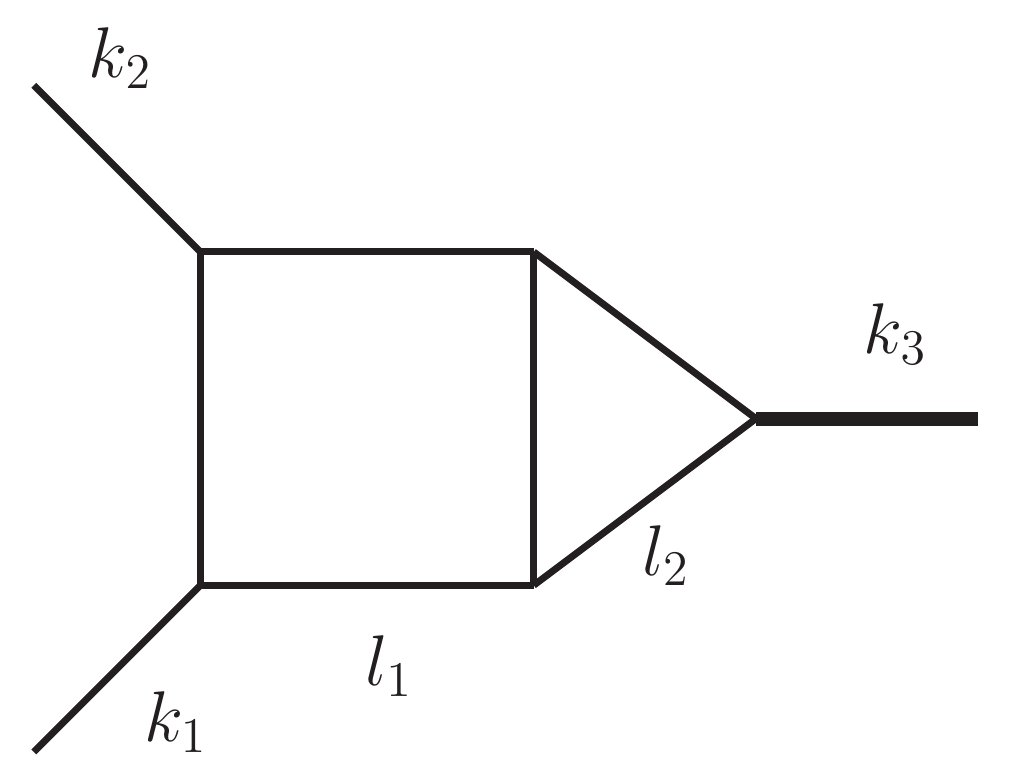}
        \caption{$m_1=m_2=0$}
        \label{fig:gull}
    \end{subfigure}
    \begin{subfigure}[b]{0.3\textwidth}
        \includegraphics[width=\textwidth]{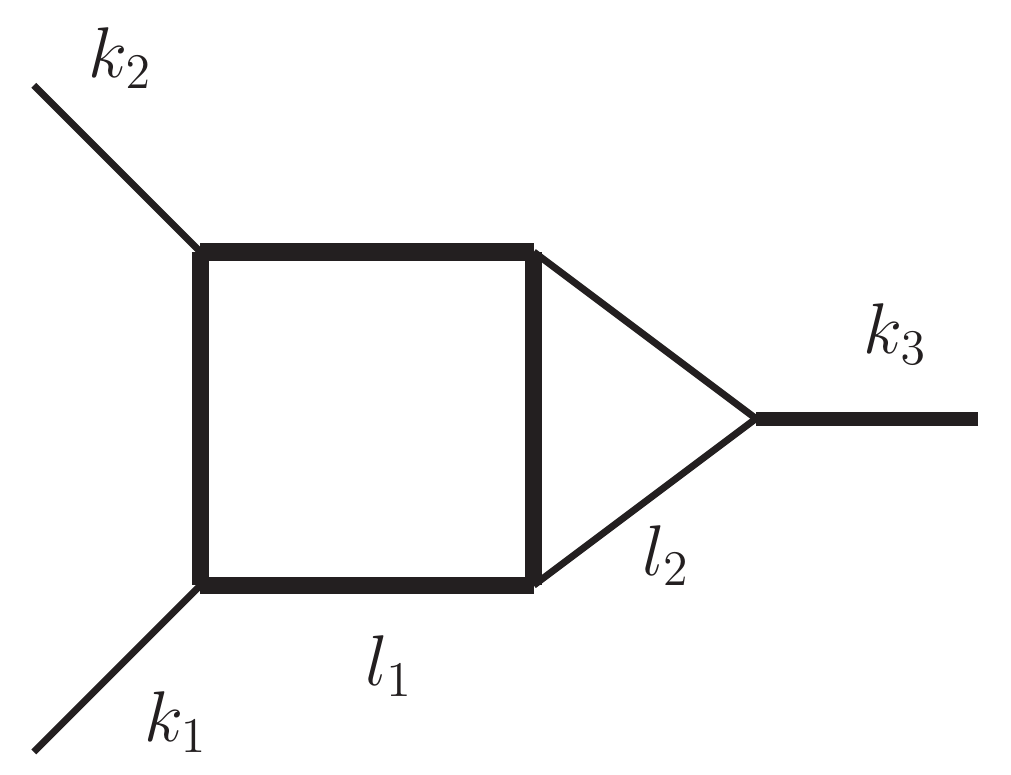}
        \caption{$m_1\not=0$, $m_2=0$}
        \label{fig:tiger}
   \end{subfigure}
    \begin{subfigure}[b]{0.3\textwidth}
        \includegraphics[width=\textwidth]{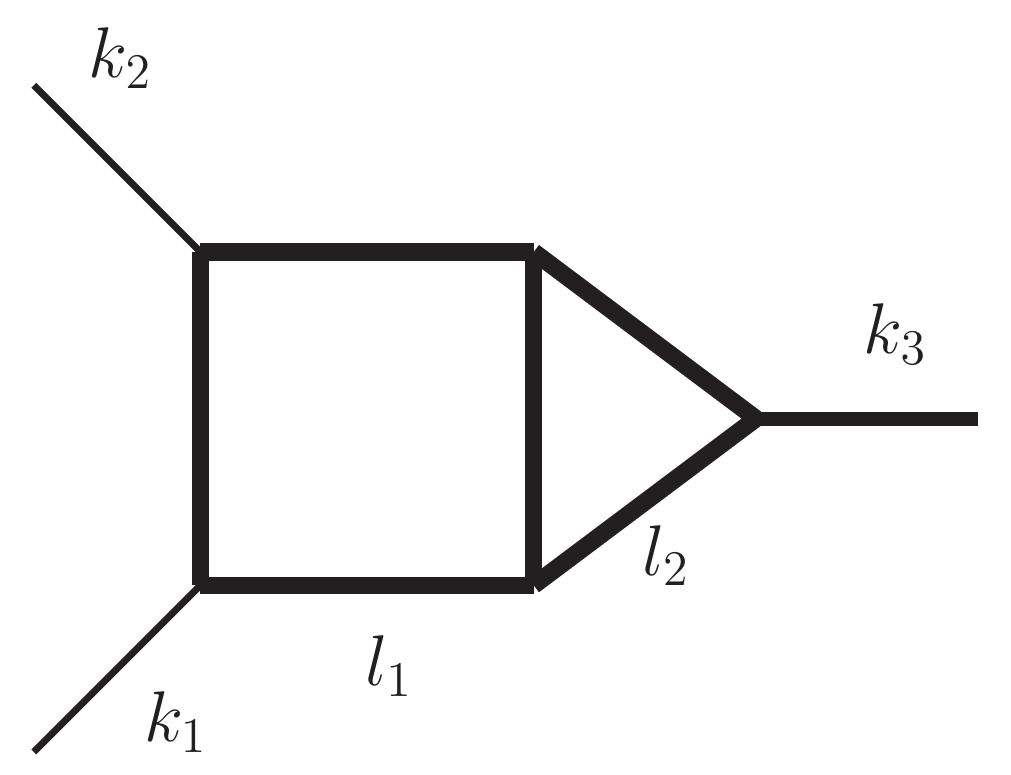}
        \caption{$m_1\not=0$, $m_2\not=0$}
        \label{fig:mouse}
    \end{subfigure}
    \caption{Two-loop box triangle diagrams}\label{box_tri}
\end{figure}

For example, we consider the $D$-dimensional two-loop three-point
box-triangle diagram, with six propagators, see figure \ref{box_tri}. The inverse propagators
are
\begin{gather}
D_1=l_1^2 - m_1^2,\quad D_2=(l_1 - k_1)^2 -
  m_1^2, \quad
D_3= (l_1 - k_1 - k_2)^2 - 
 m_1^2, \nonumber\\ D_4= (l_2 + k_1 + k_2)^2-m_2^2,\quad D_5=l_2^2-m_2^2,\quad D_6=(l_1 + l_2)^2 - m_1^2
\label{dbox_prop}
\end{gather}
and the external momenta satisfy $k_1^2=k_2^2=0$, $(k_1+k_2)^2=s$. There are $2\times
(3-1)+2\times (2+1)/2=7$ scalar products of the diagram, therefore the
number of ISPs is $7-6=1$. We may choose the ISP to be
\begin{equation}
  \label{eq:2}
  D_7=(l_2+k_1)^2\,.
\end{equation}
The Baikov variables are defined to be $z_i \equiv D_i$,
$i=1,\ldots,7$. The integrals under consideration are
\begin{equation}
  \label{eq:4}
  I[a_1, a_2, a_3, a_4, a_5, a_6,-k](D)\equiv\int \frac{d^D l_1}{\pi^{D/2}}
\frac{d^D l_2}{\pi^{D/2}} \frac{D_7^k}{D_1^{a_1} D_2^{a_2}D_3^{a_3}D_4^{a_4}D_5^{a_5}D_6^{a_6}}\,.
\end{equation}
We may drop the argument $D$ in
the notation except for the discussion of dimension shift identities. 

We consider different mass configurations:
\begin{itemize}
\item $m_1=m_2=0$. In this case, the Baikov polynomial with maximal
  cut $z_1=\ldots =z_6=0$ is
  \begin{equation}
    \label{eq:3}
    F=-\frac{1}{4} z_7^2\,.
  \end{equation}
A
generic integrand $z_7^k/(z_1 \ldots z_6)$, $k\in \mathbb N_0$, on the maximal cut reads,
\begin{equation}
  \label{boxtri_massless_mc}
 I[1,1,1,1,1,1,-k]\mc =\frac{2^{D-4}}{\pi^3 \Gamma(D-3)}\frac{1}{8
s^2}\int_\Omega dz_7 \big(-\frac{1}{4} z_7^2\big)^{(D-5)/2} z_7^k\,.
\end{equation}
Note that
\begin{equation}
  \label{eq:5}
  d\bigg(\frac{1}{D-4+k}\big(-\frac{1}{4} z_7^2\big)^{(D-5)/2}
  z_7^{k+1}\bigg)=\big(-\frac{1}{4} z_7^2\big)^{(D-5)/2} z_7^k dz_7\,,
\end{equation}
so the integrand of \eqref{boxtri_massless_mc} is a
polynomial-valued total derivative. Hence
\begin{equation}
  \label{eq:6}
   I[1,1,1,1,1,1,-k]\mc =0\,,\quad k\in\mathbb N_0\,.
\end{equation}
The $D$-dimensional maximal cut vanishes and this implies that the
massless box-triangle integral is reducible (to integrals with fewer propagators).

\item $m_1\not=0$, $m_2=0$. In this case, the Baikov polynomial with maximal
  cut $z_1=\ldots =z_6=0$ is
  \begin{equation}
    F=\frac{z_7(4m_1^2 s+4m_1^2 z_7 -s z_7)}{4s}\,.
  \end{equation}
To simplify the expression, we may redefine the ISP as
%$z_7=\frac{4 x
  %s}{4 x-1} (z-\frac{1}{2}) $ where $x=m_1^2/s$，
\begin{gather}
  \label{eq:8}
  \tilde D_7=z\equiv\frac{4m_1^2-s}{4m_1^2s}
  z_7+\frac{1}{2}=\frac{4m_1^2-s}{4m_1^2s} (l_2+p_1)^2+\frac{1}{2}\,,\\
\tilde I[a_1, a_2, a_3, a_4, a_5, a_6,-k]=\int \frac{d^D l_1}{\pi^{D/2}}
\frac{d^D l_2}{\pi^{D/2}} \frac{\tilde D_7^k}{D_1^{a_1} D_2^{a_2}D_3^{a_3}D_4^{a_4}D_5^{a_5}D_6^{a_6}}\,.
\end{gather}
Then
\begin{equation}
  \label{eq:7}
  F=\frac{4 s^2 x^2}{4 x-1} (z-\frac{1}{2}) (z+\frac{1}{2}) \,,
\end{equation}
where $x=m_1^2/s$.

We consider the kinematic region $s<4m^2$ ($x>1/4$). The
$D$-dimensional cut reads
\begin{equation}
  \label{boxtri_one_mass_mc_int}
 \tilde I[1,1,1,1,1,1,-k]\mc =\frac{2^{D-4}}{\pi^3 \Gamma(D-3)}\frac{x}{2 s
   (4x-1)}\int_\Omega dz F^{(D-5)/2} z^k\,.
\end{equation}
When $x>1/4$, $\Omega=\Omega_1\cup \Omega_2$, where
$\Omega_1=(1/2,\infty)$ and $\Omega_2=(-\infty,-1/2)$. The integration
\eqref{boxtri_one_mass_mc_int} over $\Omega_1$ is
\begin{equation}
  \label{boxtri_one_mass_mc_1}
   \tilde I[1,1,1,1,1,1,-k]\mc^{(1)} =\frac{s^{D-6}x^{D-4}
    (4x-1)^{\frac{3}{2}-\frac{D}{2}}}{2^{k+3}\pi^{\frac{5}{2}}}
  \frac{\Gamma(\frac{4-D-k}{2})}{\Gamma(\frac{D}{2}-1)
\Gamma(\frac{1}{2}-\frac{k}{2})}\,,
\end{equation}
and similarly over the region $\Omega_2$,
\begin{equation}
  \label{eq:9}
   \tilde I[1,1,1,1,1,1,-k]\mc^{(2)} = (-1)^k\tilde
I[1,1,1,1,1,1,-k]\mc^{(1)}\,.
\end{equation}
These integrals are convergent when ${\rm Re}(D)>3$ and $k+{\rm
  Re}(D)<4$. We use \eqref{boxtri_one_mass_mc_1} as an analytic
continuation for generic values of $D$ and $k$.
%The $D$-dimensional maximal cut explicitly reads,
%\begin{equation}
 % \label{boxtri_one_mass_mc}
  %I[1,1,1,1,1,1,-k]\mc =\big(1+(-1)^k\big)\frac{s^{D-6}x^{D-4}
   % (4x-1)^{\frac{3}{2}-\frac{D}{2}}}{2^{k+3}\pi^{\frac{5}{2}}}
  %\frac{\Gamma(\frac{4-D-k}{2})}{\Gamma(\frac{D}{2}-1) \Gamma(\frac{1}{2}-\frac{k}{2})}.
%\end{equation}
Note that because of the factor $1/\Gamma(\frac{1}{2}-\frac{k}{2})$, $
\tilde I[1,1,1,1,1,1,-k]\mc^{(1)}=0$ for positive odd $k$. So for $k\in N_0$,
\begin{equation}
  \label{eq:10}
 \tilde I[1,1,1,1,1,1,-k] \mc^{(1)}=\tilde I[1,1,1,1,1,1,-k]\mc^{(2)}\,,\quad
k\in N_0\,.
\end{equation}
Hence we define the $D$-dimensional maximal cut as 
\begin{equation}
  \label{boxtri_one_mass_mc}
  \tilde  I[1,1,1,1,1,1,-k]\mc \equiv \tilde I[1,1,1,1,1,1,-k] \mc^{(1)}\,.
\end{equation}

By expression \eqref{boxtri_one_mass_mc}, we find that
\begin{eqnarray}
  \label{eq:11}
 \tilde I[1,1,1,1,1,1,-2] \mc&=&\frac{1}{4(D-2)} \tilde I[1,1,1,1,1,1,0] \mc \,,\\
\tilde I[1,1,1,1,1,1,-4] \mc&=&\frac{3}{16D(D-2)} \tilde I[1,1,1,1,1,1,0] \mc \,,\\
\tilde I[1,1,1,1,1,1,-6] \mc&=&\frac{15}{64(D+2)D(D-2)} \tilde I[1,1,1,1,1,1,0] \mc\,.
\end{eqnarray}
They imply the Feynman integral relations, 
\begin{eqnarray}
  \label{eq:11}
 \tilde I[1,1,1,1,1,1,-2] &=&\frac{1}{4(D-2)} \tilde I[1,1,1,1,1,1,0]
                              +\ldots \,,\\
 \tilde I[1,1,1,1,1,1,-4] &=&\frac{3}{16D(D-2)} \tilde
                              I[1,1,1,1,1,1,0] +\ldots \,,\\
 \tilde I[1,1,1,1,1,1,-6] &=&\frac{15}{64(D+2)D(D-2)} \tilde
                              I[1,1,1,1,1,1,0] + \ldots \,,
\end{eqnarray}
where $\ldots$ stands for integrals with fewer propagators. These
identities agree with IBP output from {\rm FIRE}
\cite{Smirnov:2005ky,Smirnov:2006tz, Smirnov:2008iw, Smirnov:2014hma},
{\rm LiteRed} \cite{Lee:2012cn,Lee:2013mka},
and {\rm Azurite}  \cite{Georgoudis:2016wff}.

The expression \eqref{boxtri_one_mass_mc} can also be used for
deriving the dimension shift identity,
\begin{equation}
  \label{eq:12}
  \tilde I[1,1,1,1,1,1,0](D+2) \mc = -\frac{4 m^4 s}{(D-2)^2 (4m^2-s)} \tilde I[1,1,1,1,1,1,0](D) \mc \,,
\end{equation}
which means the dimension shift identity on the maximal cut.

It is also interesting to study the differential equation on the maximal cut,
\begin{equation}
  \label{DE_boxtri_massive}
  \frac{\partial }{\partial x} \tilde I[1,1,1,1,1,1,0]
  =\bigg(-2\frac{D-3}{4x-1}+\frac{D-4}{x}\bigg) \tilde I[1,1,1,1,1,1,0]
  +\ldots\,,
\end{equation}
where $\ldots$ stands for integrals with fewer
propagators. Explicitly, we see that the $D$-dimensional cut
$\tilde  I[1,1,1,1,1,1,0]\mc$ solves the maximal cut part of this equation.
%\begin{equation}
 % \label{DE_boxtri_massive}
 % \frac{\partial }{\partial x} I[1,1,1,1,1,1,0]\mc
 % =\bigg(-2\frac{D-3}{4x-1}+\frac{D-4}{x}\bigg)  I[1,1,1,1,1,1,0]\mc,
%\end{equation}

We can expand the $D$-dimensional maximal cut in $\epsilon$ $(D=4-2\epsilon)$,
\begin{gather}
  \label{eq:14}
   \tilde I[1,1,1,1,1,1,0]\mc=\frac{1}{8 \pi ^3 s^2 \sqrt{4 x-1} \epsilon
    }+\frac{-2 \log (s)-2 \log (x)+\log (4 x-1)-2 \gamma }{8 \pi ^3
      s^2 \sqrt{4 x-1}}
\nonumber\\+\frac{\epsilon  }{16 \pi ^3 s^2 \sqrt{4 x-1}}\bigg(8 \log (s) \log
(x)-4 \log (s) \log (4 x-1)+4 \log ^2(s)+8 \gamma  \log (s)+4 \log
^2(x)\nonumber\\+\log ^2(4 x-1)+8 \gamma  \log (x)-4 \log (x) \log (4 x-1)-4
\gamma  \log (4 x-1)+4 \gamma ^2\bigg)+O\left(\epsilon ^2\right)\,,
\end{gather}
where $\gamma$ is the Euler-Mascheroni constant.
From the leading coefficient in $\epsilon$, we see that we may redefine the integral,
\begin{equation}
  \label{eq:18}
  J\equiv s^2\sqrt{4x-1}\tilde I[1,1,1,1,1,1,0]\,,
\end{equation}
such that the differential equation on the maximal cut has the
$\epsilon$ form,
\begin{equation}
  \label{DE_canonical_boxtri_massive}
  \frac{\partial }{\partial x} J
  =\epsilon\frac{2-4x}{(4x-1)x}
  J+\ldots ,
\end{equation} 
where $\ldots$ stands for integrals with fewer propagators. 

\item $m_1\not=0$, $m_2\not=0$. Define $x_1=m_1^2/s$ and
  $x_2=m_2^2/s$. Following similar steps as in the previous case, we
  define
\begin{gather}
  \tilde D_7=z\equiv\frac{4m_1^2-s}{4m_1^2s}
  z_7+\frac{1}{2}-\frac{m_2^2}{s}=\frac{4m_1^2-s}{4m_1^2s} (l_2+p_1)^2+\frac{1}{2}-\frac{m_2^2}{s}\,,\\
\tilde I[a_1, a_2, a_3, a_4, a_5, a_6,-k]=\int \frac{d^D l_1}{\pi^{D/2}}
\frac{d^D l_2}{\pi^{D/2}} \frac{\tilde D_7^k}{D_1^{a_1} D_2^{a_2}D_3^{a_3}D_4^{a_4}D_5^{a_5}D_6^{a_6}}\,.
\end{gather}
We find that the $D$-dimensional maximal cut is
\begin{eqnarray}
  \label{eq:15}
  &&\tilde I[a_1, a_2, a_3, a_4, a_5, a_6,-k]\mc\nonumber\\
&&=\frac{s^{D-6}x_1^{D-4}
    (4x_1-1)^{\frac{3}{2}-\frac{D}{2}}(1-4x_2+\frac{x_2^2}{x_1})^\frac{D+k-4}{2}}{2^{k+2}\pi^{\frac{5}{2}}}
  \frac{\Gamma(\frac{4-D-k}{2})}{\Gamma(\frac{D}{2}-1)
\Gamma(\frac{1}{2}-\frac{k}{2})}\,.
\end{eqnarray}
Again, we can use it to study the IBPs, dimension shift identities,
differential equations and $\epsilon$-form, on the maximal cut. %For example, 
%\begin{align}
 %\label{eq:16}
 % \tilde I[a_1, a_2, a_3, a_4, a_5, a_6,-2]
 % &=\frac{m_2^4+m_1^2s-4m_1^2 m_2^2}{4(D-2)m_1^2s}\tilde I[a_1, a_2, a_3,
 % a_4, a_5, a_6,0]+\ldots\nonumber\\
 % \tilde I[a_1, a_2, a_3, a_4, a_5, a_6,-4]
 % &=\frac{3(m_2^4+m_1^2s-4m_1^2 m_2^2)^2}{16(D-2)D m_1^4s^2}\tilde I[a_1, a_2, a_3, a_4, a_5, a_6,%0]+\ldots
%\end{align}

 %These
%identities agree with IBP output from {\rm FIRE}
%\cite{Smirnov:2005ky,Smirnov:2006tz, Smirnov:2008iw, Smirnov:2014hma}
%and {\rm Azurite}  \cite{Georgoudis:2016wff}.

\end{itemize}

%%%%%%%%%%%%%%%%%%%%%
\section{Planar maximal cuts with two ISPs}
%%%%%%%%%%%%%%%%%%%%%%%
Now we consider $D$-dimensional maximal cuts with two
ISPs. In this case, the integral regions become less trivial and we
can investigate each region seperately to get a complete set of $D$-dimensional
maximal cut functions in kinematic variables.

\subsection{Massless sunset}
\begin{figure}[h!]
\centering
\includegraphics[scale=1]{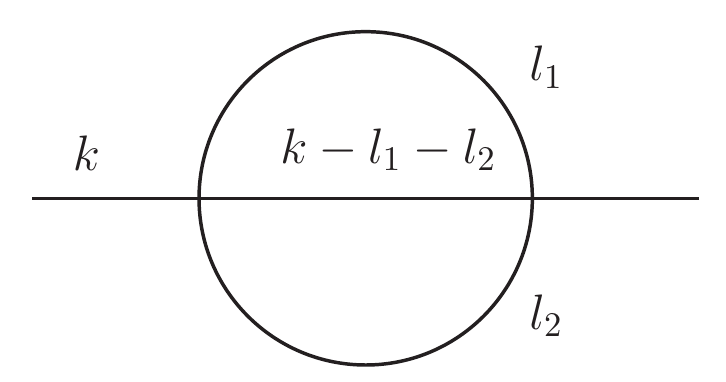}
\caption[Caption]{The massless sunset diagram.\footnotemark All momenta flow to the right.}
\label{fig:sunset}
\end{figure}
\begin{figure}
\centering
\includegraphics[scale=0.7]{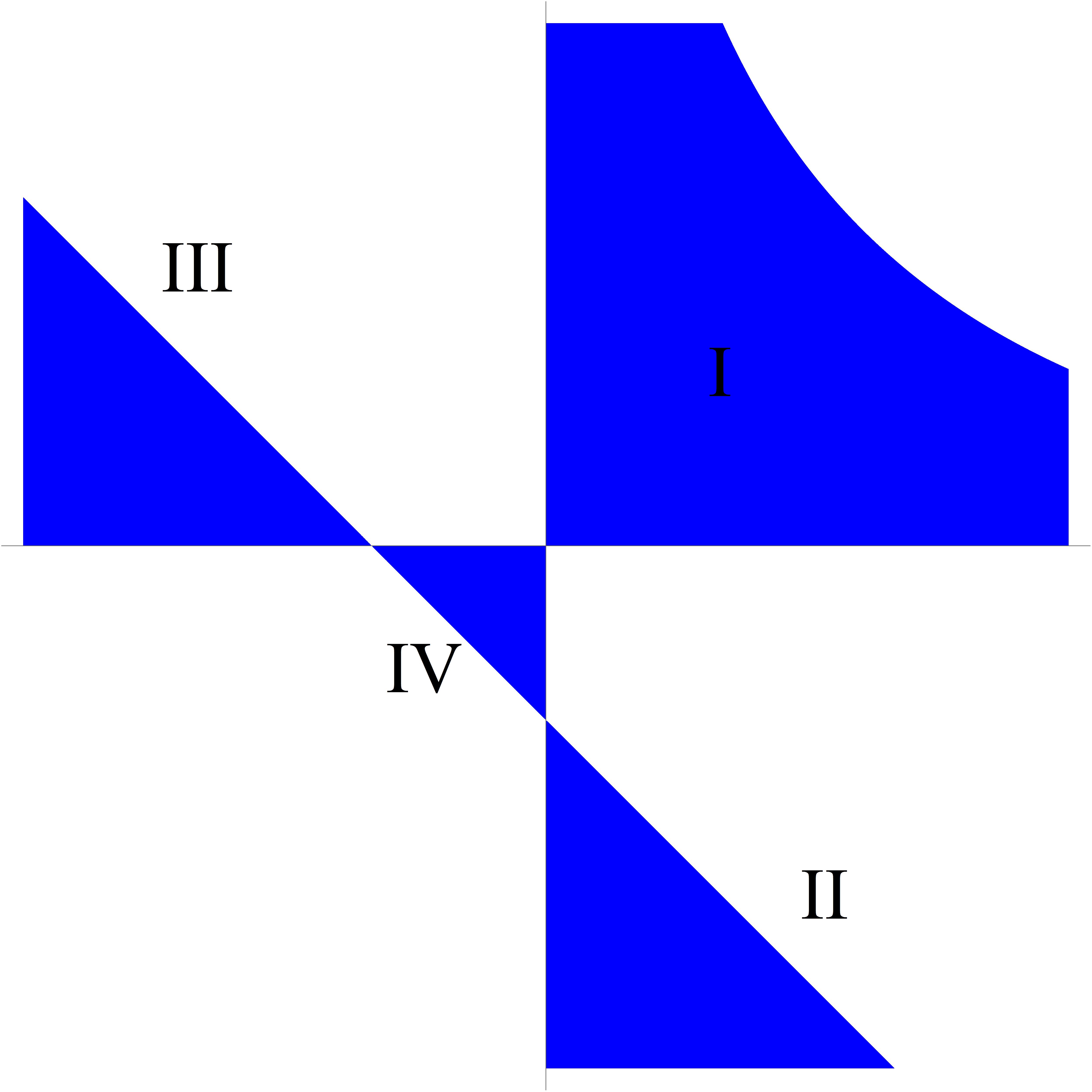}
\caption{Subregions for the integration of the sunset
  diagram on the maximal cut.}
\label{fig:sunsetregions}
\end{figure}

\footnotetext{The diagram was drawn using JaxoDraw \cite{Vermaseren:1994je,Binosi:2008ig}.}
As a warm-up, consider the sunset diagram with massless internal propagators, see figure \ref{fig:sunset}.
We have the inverse propagators
\begin{gather}
	D_1 = l_1^2\,, \quad D_2 = l_2^2\,, \quad D_3 = (l_1+l_2-k)^2\,.
\end{gather}
There are two ISPs, which we may choose to be
\begin{gather}
	D_4 = (l_1 + k)^2 - 2 s\,, \quad D_5 = (l_2+k)^2 - 2s\,,
\end{gather}
where $s=k^2$.
Our object of interest is the integral\begin{equation}\label{eq:sunset1}
	I[a_1,a_2,a_3,-a_4,-a_5] \equiv \int\limits \frac{d^Dl_1}{\pi ^{D/2}}
\frac{d^Dl_2}{\pi ^{D/2}}
\frac{D_4^{a_4}D_5^{a_5}}{D_1^{a_1}D_2^{a_2}D_3^{a_3}}\,.
\end{equation}
Taking the Baikov variables to be $z_i = D_i$, equation \eqref{eq:sunset1} becomes
\begin{equation}\label{eq:sunset2}
	I[a_1,a_2,a_3,-a_4,-a_5] = \frac{1}{s} \frac{2^{D-4}
\pi^{D-2}}{\Gamma(D-2)} \int\limits_\Omega \left( \prod\limits_{i=1}^5 dz_i
\right) F^{ \frac{D-4}{2} }
\frac{z_4^{a_4}z_5^{a_5}}{z_1^{a_1}z_2^{a_2}z_3^{a_3}}\,,
\end{equation}
where the integration region $\Omega$ is defined by $F(z)\geq 0$.
On the maximal cut the Baikov polynomial is 
\begin{equation}
	F = \frac{1}{4 s} z_4 z_5 (s + z_4 + z_5)\,,
\end{equation}
which we can simplify by rescaling the Baikov variables to
\begin{gather}
	x = \frac{z_4}{s}\,, \quad y = \frac{z_5}{s}\,.
\end{gather}
This gives\begin{equation}
	F = xy(1+x+y)\,,
\end{equation}
and our integral of interest on the maximal cut then reads \begin{equation}
	J[a,b]\mc^{(\Omega)} = 
	s^2 \left(\frac{s^2}{4}\right)^{ \frac{D-4}{2}} \frac{2^{D-4}
\pi^{D-2}}{\Gamma(D-2)} s^{a+b} \int\limits_\Omega F^{ \frac{D-4}{2} }
x^{a}y^{b}dxdy\,,
\end{equation}
where we have defined the shorthand $J[a,b] \equiv I[1,1,1,-a,-b]$ for notational convenience.
The integration splits into four regions,
\begin{itemize}
  \item[$\Omega_\text{I}$:] $x,y > 0$,
  \item[$\Omega_\text{II}$:] $x > 0$, $y < -(1+x)$,
  \item[$\Omega_\text{III}$:] $x < -(1+y)$, $y > 0$,
  \item[$\Omega_\text{IV}$:] $-1 < x < 0$, $-(1+x) < y < 0$.
\end{itemize}
These regions are shown schematically in figure \ref{fig:sunsetregions}.

The integration in each of the regions can now be carried out explicitly to give
%\begin{gather}
\begin{align*}
	I_\text{I} & = \int\limits_0^\infty \int\limits_0^\infty x^a y^b (xy(1+x+y))^{ \frac{D-4}{2}} dydx\\
		& = \frac{\Gamma(4-a-b- 3D/2) \Gamma(-1+a+D/2)\Gamma(-1+b+D/2)}{\Gamma(2-D/2)}\,,\displaybreak[0]\\
	I_\text{II} & = \int\limits_0^\infty \int\limits_{-\infty}^{-(1+x)} x^a y^b (xy(1+x+y))^{ \frac{D-4}{2}} dydx\\
		& = (-1)^b \frac{\Gamma(4-a-b-3D/2) \Gamma(-1+a+D/2)\Gamma(-1+D/2)}{\Gamma(2-b-D/2)}\,,\displaybreak[0]\\
	I_\text{III} & = \int\limits_0^\infty \int\limits_{-\infty}^{-(1+y)} x^a y^b (xy(1+x+y))^{ \frac{D-4}{2}} dxdy\\
		& = (-1)^a \frac{\Gamma(4-a-b-3D/2) \Gamma(-1+D/2) \Gamma(-1+b+D/2)}{\Gamma(2-a-D/2)}\,,\displaybreak[0]\\
	I_\text{IV} & = \int\limits_{-1}^0 \int\limits_{-(1+x)}^0 x^a y^b (xy(1+x+y))^{ \frac{D-4}{2}} dydx\\
		& = \frac{(-1)^{1+a+b} \pi}{\sin(D\pi/2)} \,\frac{\Gamma(-1+a+D/2)\Gamma(-1+b+D/2)}{\Gamma(2-D/2)\Gamma(-3+a+b+3D/2)}\,.
\end{align*}
%\end{gather}
Note that the integrals only converge when certain restrictions on $a$, $b$, and $D$ are satisfied,
but we drop those as we are interested in the analytic continuation anyway.
Assuming $a$ and $b$ to be non-negative integers,
it follows from the reflection property of the gamma function,
$\sin(\pi z) \Gamma(1-z)\Gamma(z) = \pi$, that
\begin{equation}
	I_\text{I}=I_\text{II}=I_\text{III}\,, \quad I_\text{IV} = (1+2\cos(D\pi))I_\text{I}\,.
\end{equation}
So there is only one linearly independent function. (Here the
coefficient field is set to be the meromorphic function field of $D$.) 
We thus set the $D$-dimensional maximal cut as the integration over
the region I,
\begin{gather}\begin{aligned}
	J[a,b]\mc & = 
	\frac{\pi^{D-2}}{\Gamma(D-2)} s^{a+b+D-2} \\
	& \quad \times \frac{\Gamma(4-a-b- 3D/2) \Gamma(-1+a+D/2)\Gamma(-1+b+D/2)}{\Gamma(2-D/2)}\,.
\end{aligned}\end{gather}
If we introduce the descending factorial
\begin{equation}
	(z)^{(\underline{r})} \equiv \frac{\Gamma(z+1)}{\Gamma(z-r+1)} = z(z-1)\cdots(z-r+1),
\end{equation}
which is related to the Pochhammer symbol \eqref{eq:Pochhammer} by
\begin{equation}
	(z)_r = (-1)^r (-z)^{(\underline{r})}\,,
\end{equation}
it is straightforward to see that
\begin{equation}\label{eq:sunset_generalIBP}
	J[a,b]\mc = J[0,0]\mc \, s^{a+b} \frac{ \left(-1 +
\frac{D}{2}\right)_{a} \left(-1+ \frac{D}{2}\right)_{b} }{\left(3-\frac{3D}{2}\right)^{(\underline{a+b})}}\,.
\end{equation}
For $a$ and $b$ non-negative integers the last factor evaluates to a rational function in $D$,
as one would expect.
Equation \eqref{eq:sunset_generalIBP} agrees with IBPs found with {\rm FIRE}
\cite{Smirnov:2005ky,Smirnov:2006tz, Smirnov:2008iw, Smirnov:2014hma},
{\rm LiteRed} \cite{Lee:2012cn,Lee:2013mka},
and {\rm Azurite}  \cite{Georgoudis:2016wff}.

In a similar way we readily find the dimension shift identity,
\begin{gather}\begin{aligned}
	J[0,0]\mc(D+2)  = - \, s^2 \pi^2 
		  \frac{ (-1+D/2)(-1+D/2) }{2(1-3D/2)_{3}(D-1)} \times J[0,0]
                  \mc(D)\,.
\end{aligned}\end{gather}
The differential equation for $J[0,0]\mc$ is simple,
\begin{equation}
	\frac{\partial}{\partial s} J[0,0]\mc = \frac{D-2}{s} J[0,0]\mc\,,
\end{equation}
and is immediately in $\epsilon$-form if we specialize to $D=2-2\epsilon$.

\subsection{Massless double box}
\begin{figure}[h]
  \centering
  \includegraphics{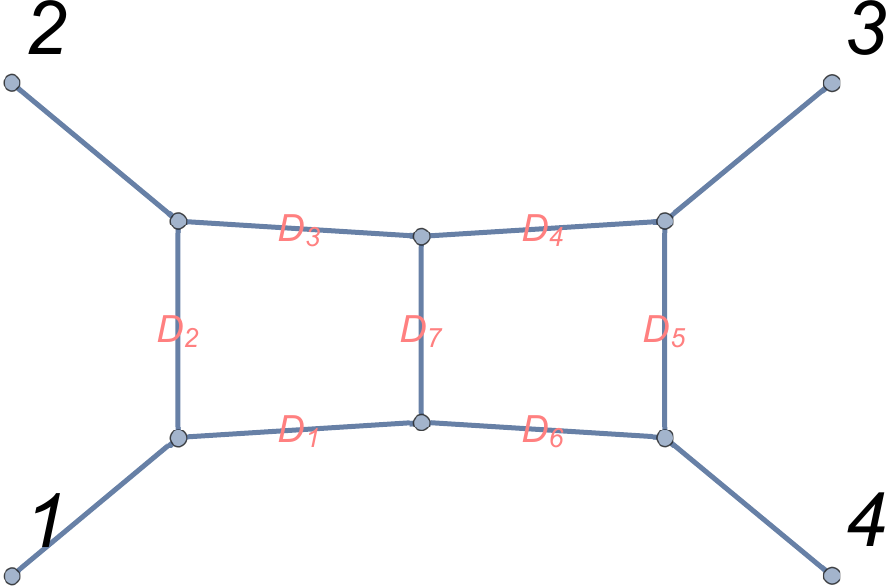}
\caption{Massless double box diagram. This graph is produced by the package {\sc
  Azurite} \cite{Georgoudis:2016wff}.}
\label{dbox_graph}
\end{figure}

The diagram is shown in figure \ref{dbox_graph}.
The inverse propagators
are
\begin{gather}
  \label{eq:1}
D_1=l_1^2\,,\quad D_2=(l_1 - k_1)^2\,, \quad
D_3= (l_1 - k_1 - k_2)^2\,, \nonumber\\ 
D_4= (l_2 + k_1 + k_2)^2\,,\quad D_5=(l_2-k_4)^2\,,\quad D_6= l_2^2\,, 
\quad D_7=(l_1+l_2)^2\,,
\end{gather}
and the external momenta satisfy $k_1^2=k_2^2=k_4^2=0$,
$(k_1+k_2)^2=s$ and $(k_1+k_4)^2=t$. We define $\chi=t/s$. There are
$2(4-1)+2(2+1)/2=9$ scalar products in loop momenta, hence we have
two ISPs,
\begin{equation}
  \label{eq:19}
  D_8=(l_1+k_4)^2\,,\quad D_9=(l_2+k_1)^2\,.
\end{equation}
The integrals under consideration are
\begin{equation}
  \label{eq:21}
  I[n_1,n_2,\ldots, n_7,-a,-b]\equiv \int \frac{d^D
    l_1}{\pi^{D/2}}\frac{d^D l_2}{\pi^{D/2}} \frac{D_8^a D_9^b}{D_1^{n_1}
    D_2^{n_2} \ldots D_7^{n_7}}\,.
\end{equation}
To simplify notation, define $J[a,b]\equiv
I[1,1,1,1,1,1,1,-a,-b]$. Again we hide the argument $D$, except for
the discussion of dimensional shift identities.

The Baikov polynomial on the maximal cut is
\begin{equation}
  \label{eq:20}
  F=\frac{z_8 z_9 \left(s^2 \chi -s z_8-s z_9-z_8 z_9\right)}{4 s^2 \chi  (\chi +1)}\,.
\end{equation}

\begin{figure}[h!]
\centering
\includegraphics[scale=0.7]{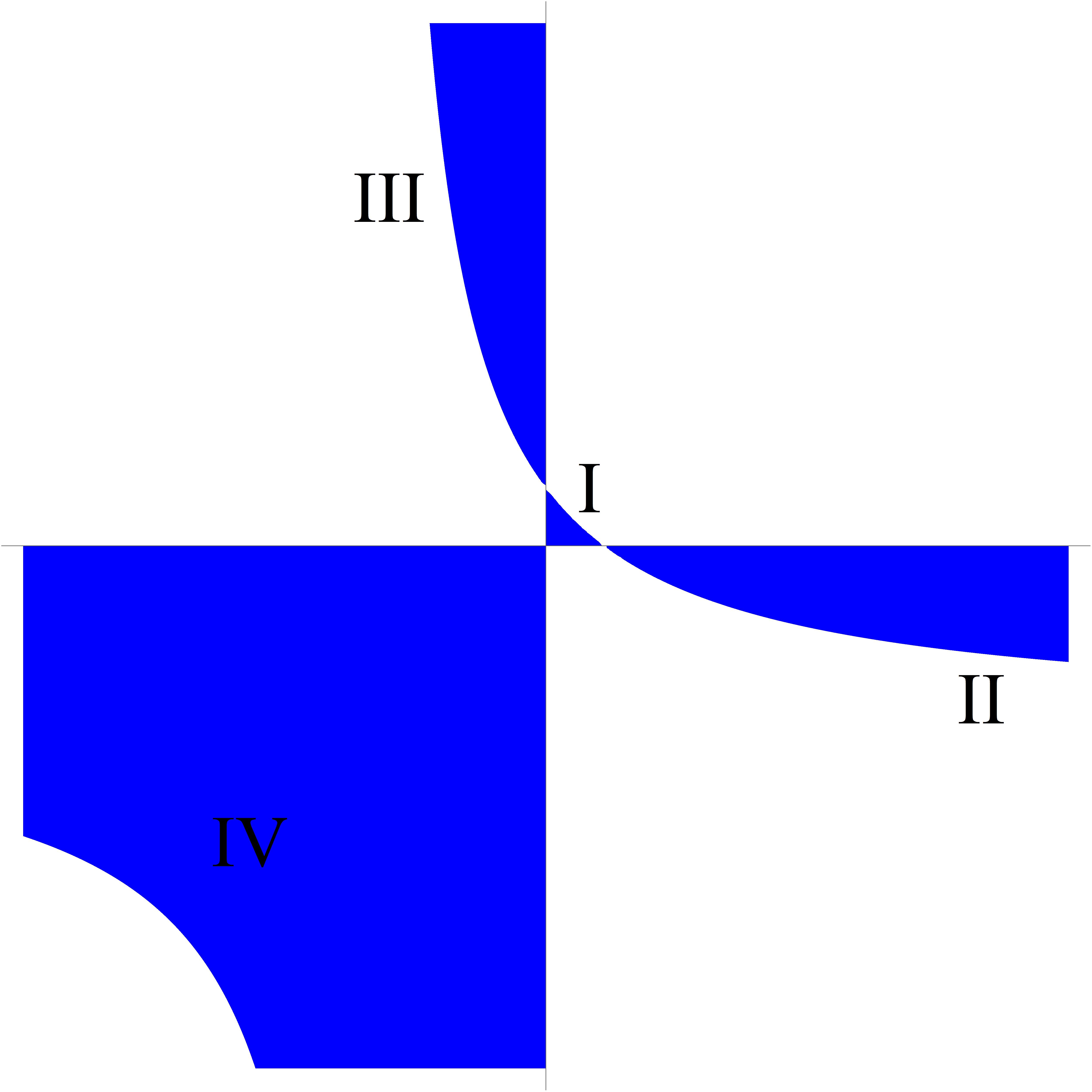}
\caption{Subregions for the integration of the massless double box
  diagram on the maximal cut. This plot is for the kinematic
  configuration, $s=1$ and $\chi=1/3$. The two axes are the ISPs,
  $z_8$ and $z_9$. These subregions are symmetric under the
flip of two axes.}
\label{dbox_region}
\end{figure}

The maximal cut of $J[a,b]$ can be calculated by the integration of
$F^{(D-6)/2} z_8^a z_9^b$. Consider the kinematic condition $s>0,
\chi>0$. The integration region defined by $F\geq 0$ splits into four
subregions:
 \begin{itemize}
  \item[$\Omega_\text{I}$:] $z_8>0$, $z_9>0$, $s^2 \chi- s z_8- s z_9- z_8
    z_9>0$,
  \item[$\Omega_\text{II}$:] $z_8>0$, $z_9<0$, $s^2 \chi- s z_8- s z_9- z_8
    z_9<0$,
 \item[$\Omega_\text{III}$:] $z_8<0$, $z_9>0$, $s^2 \chi- s z_8- s z_9- z_8
    z_9<0$,
  \item[$\Omega_\text{IV}$:] $z_8<0$, $z_9<0$, $s^2 \chi- s z_8- s z_9- z_8
    z_9>0$,
 \end{itemize}
which are shown in figure \ref{dbox_region} as the blue area. It is
clear that on the four subregions the conditions
$\mu_{11}(z)\geq 0$ and $\mu_{22}(z)\geq 0$ are satisfied.

The integrations over the first three subregions are straightforward,
while the integration over the fourth subregion needs careful further
splitting. The integration over $\Omega_\text{I}$ reads,
\begin{align}
  \label{eq:22}
  J[a,b]\mc ^\text{(I)}&=\frac{2^{D-10}}{\pi ^4 s t \Gamma (D-4)
                        (s+t)}\int_0^{\chi s} dz_8
                        \int_0^{\frac{s(\chi s-z_8)}{s+z_8}} dz_9 F^{\frac{D-6}{2}} z_8^a z_9^b\nonumber\\
&=\frac{\Gamma \left(\frac{D}{2}-2\right) \Gamma
    \left(a+\frac{D}{2}-2\right) \Gamma \left(b+\frac{D}{2}-2\right)
    s^{a+b+D-7} \chi ^{a+b+D-5} \,}{16 \pi ^4 \Gamma (D-4)}\nonumber\\
 &\times \ _2\tilde{F}_1\left(a+D-4,b+D-4;a+b+\frac{3 D}{2}-6;-\chi \right)\,,
\end{align}
where $_2\tilde{F}_1$ is the regularized hypergeometric function,
$_2\tilde{F}_1(\alpha,\beta,\gamma,z)=\,_2F_1(\alpha,\beta,\gamma,z)/\Gamma(\gamma)$. This
result is to be understood as an analytic continuation over $D$, $a$
and $b$.

Similarly the second integration over $\Omega_\text{II}$ gives
\begin{align}
  \label{eq:17}
 J[a,b]\mc ^\text{(II)}& = \frac{(-1)^b 
                        \Gamma \left(\frac{D}{2}-2\right) \Gamma (-a-D+5) \Gamma
                        \left(b+\frac{D}{2}-2\right) s^{a+b+D-7} \chi
                        ^{a+\frac{D}{2}-3} (\chi +1)^{2-\frac{D}{2}}}{16
                        \pi ^4 \Gamma (D-4)}\nonumber \\
& \times \, _2\tilde{F}_1\left(-a-D+5,b+\frac{D}{2}-2;-a+b+1;-\frac{1}{\chi}\right)\,.
\end{align}
Although apparently this expression does not look symmetric in $a$ and
$b$, after a hypergeometric function transformation, the symmetry is manifest:
\begin{align}
  \label{eq:18}
 J[a,b]\mc ^\text{(II)}& =-\frac{(-1)^{a+b}\chi ^{2-\frac{D}{2}} \sin
                        (\pi  D) \Gamma \left(\frac{D}{2}-2\right)
                        \Gamma (-a-D+5) \Gamma (-b-D+5)
                        s^{a+b+D-7}}{16 \pi ^4 \Gamma (D-4) \sin
                        \left(\frac{3 \pi  D}{2}\right)}\nonumber\\
&\times  \,
                        _2\tilde{F}_1\left(-a-\frac{D}{2}+3,-b-\frac{D}{2}+3;-a-b-\frac{3
                        D}{2}+8;-\chi \right)\nonumber\\
&+\frac{\Gamma \left(\frac{D}{2}-2\right) \Gamma
    \left(a+\frac{D}{2}-2\right) \Gamma \left(b+\frac{D}{2}-2\right)
    s^{a+b+D-7} \chi ^{a+b+D-5} \,}{16 \pi ^4 \Gamma (D-4)
 \big (1+2\cos(\pi D) \big)}\nonumber\\
 &\times \ _2\tilde{F}_1\left(a+D-4,b+D-4;a+b+\frac{3 D}{2}-6;-\chi \right)\,.
\end{align}
The integration over $\Omega_\text{III}$ gives
\begin{equation}
  \label{eq:23}
   J[a,b]\mc ^\text{(III)} = J[a,b]\mc ^\text{(II)}\,.
\end{equation}
Finally, using the transformation identities of hypergeometric functions, the integration over $\Omega_\text{IV}$ is
\begin{align}
  \label{dbox_IV}
   J[a,b]\mc ^\text{(IV)} =  J[a,b]\mc ^\text{(I)} -2  \cos(\pi D ) J[a,b]\mc^\text{(II)}\,.
\end{align}
Since the integrations over region III and IV are dependent of the
integrations over the first two regions, we can define the
$D$-dimensional maximal
cut as a list of two functions,
\begin{align}
  \label{dbox_mc}
  J[a,b]\mc \equiv \bigg( J[a,b]\mc ^\text{(I)},\  J[a,b]\mc ^\text{(II)}\bigg)\,.
\end{align}
The independence of $J[a,b]\mc ^\text{(I)}$ and $J[a,b]\mc
^\text{(II)}$ will be discussed later on. 

Then by Gauss' contiguous relations of $\,_2F_1$ functions, we see
that for integer-valued $a$ and $b$'s, $J[a,b]\mc ^\text{(I)}$'s are linearly generated by $J[0,0]\mc
^\text{(I)}$, $J[1,0]\mc ^\text{(I)}$, in the field of rational
functions of $D$, $s$ and $\chi$. Similarly, $J[a,b]\mc ^\text{(II)}$'s are linearly generated by $J[0,0]\mc
^\text{(II)}$, $J[1,0]\mc ^\text{(II)}$. Explicitly, we can check that
Gauss' contiguous relations of 
all regional integrations,  $J[a,b]\mc^{(i)}$, $i=$I, II, III, IV, provide the
maximal cut $D$-dimensional IBPs,
\begin{align}
  \label{eq:26}
  J[1,1]\mc^{(i)}& =\frac{1}{2}s^2 J[0,0]\mc^{(i)} -\frac{3}{2} s
  J[1,0]\mc^{(i)}\,, \nonumber \\
J[2,0]\mc^{(i)}&=\frac{(D-4) s^2 \chi  }{2 (D-3)}J[0,0]\mc^{(i)}
                 -\frac{ (3 D -2  \chi -12 )s}{2 (D-3)}
                 J[1,0]\mc^{(i)}\,,\nonumber\\
J[2,1]\mc^{(i)}&=
\frac{s^2 (2 D  \chi +9 D -8  \chi -30 )}{4 (D-3)}
  J[1,0]\mc^{(i)}-\frac{(3 D-10) s^3 \chi  }{4 (D-3)}
                 J[0,0]\mc^{(i)}\,.
\end{align}
These
identities agree with the IBP output from {\rm FIRE}
\cite{Smirnov:2005ky,Smirnov:2006tz, Smirnov:2008iw, Smirnov:2014hma},
{\rm LiteRed} \cite{Lee:2012cn,Lee:2013mka},
and {\rm Azurite}  \cite{Georgoudis:2016wff}. We conclude that the function relations for  $J[a,b]\mc$ provide
the IBPs on the maximal cuts. 

Gauss' contiguous relations also imply dimension shift identities
on the maximal cut. For example,  for $i=$I, II, III, IV, 
\begin{gather}
  \label{eq:27}
  J[0,0]\mc^{(i)}(D+2)=
\frac{s^2 (2 (D-3) \chi +3 d-10)}{8 (D-3)^3 (\chi
  +1)}J[0,0]\mc^{(i)}(D)\nonumber\\
-\frac{s (8 D \chi +9 D-26 \chi -30)}{8 (D-3)^3 \chi  (\chi +1)}J[1,0]\mc^{(i)}(D)\,.
\end{gather}
Note that the $\cos(\pi D)$ factor in the region \eqref{dbox_IV} does
not affect dimension shift identities, since it is invariant under
$D\to D+2$. We also remark that it is well-known that the
solutions of recursive relations for the dimension shift identities
may consist of hypergeometric functions \cite{Lee:2012te}.

We know that the two master integrals of the double box topology
satisfy the differential equation on the maximal cut,
\begin{equation}
  \label{DE_dbox}
  \frac{\partial}{\partial \chi} \left(\begin{array}{c}
J[0,0]\\
J[1,0]
\end{array}\right)=
\left(
\begin{array}{cc}
 \frac{D-\chi -5}{\chi  (\chi +1)} & \frac{D-4}{s \chi  (\chi +1)} \\
 \frac{(D-4) s}{2 (\chi +1)} & -\frac{D-4}{2 \chi  (\chi +1)} \\
\end{array}
\right) \left(\begin{array}{c}
J[0,0]\\
J[1,0]
\end{array}\right) +\ldots \,,
\end{equation}
where $\ldots $ stands for integrals with fewer propagators. We 
verify that for $i=$ I, II, III, IV,  $(J[0,0]\mc^{(i)},
J[1,0]\mc^{(i)})^{\rm T}$ solves \eqref{DE_dbox}. The maximal cut functions $J[a,b]\mc$
with $(a,b)=(0,0)$, $(1,0)$ form the fundamental solutions of
\eqref{DE_dbox},
\begin{equation}
  \label{eq:29}
  S= \left( 
    \begin{array}{c}
      J[0,0]\mc\nonumber\\
J[1,0]\mc
    \end{array}
\right) \,,
\end{equation}
because the Wronskian is nonzero,
\begin{equation}
  \label{eq:24}
  \det S \propto \chi^{\frac{d}{2}-3} (\chi +1)^{2-\frac{d}{2}} \,.
\end{equation}
This justifies the definition of \eqref{dbox_mc} since $J[a,b]\mc^\text{I}$
and $J[a,b]\mc^\text{II}$ are independent functions in $\chi$. 

Finally we check the matrix $S$ in the $D\to 4$ limit and extract the
leading coefficients in $\epsilon$. Using the {\sc HypExp} package \cite{Huber:2005yg,Huber:2007dx}, the result is
\begin{equation}
  \label{eq:28}
  T=\left(
\begin{array}{cc}
 \frac{3}{4 \pi ^4 s^3 \chi } & \frac{1}{4 \pi ^4 s^3 \chi } \\
 \frac{1}{4 \pi ^4 s^2} & 0 \\
\end{array}
\right)\,.
\end{equation}
From the leading coefficients in $T$ and a suitable linear combination, we may
define the transformation matrix
\begin{align}
  \label{eq:46}
  \tilde T=T \left(\begin{array}{cc}
 0 & 1 \\
 1 & -3
\end{array}
\right) \cdot 4 \pi^4\,.
\end{align}
Hence we redefine the master integrals as
\begin{equation}
  \label{eq:30}
  \tilde{J}[0,0] = s^3 \chi J[0,0]\,, \quad    \tilde{J}[1,0]= s^2 \chi J[1,0]\,,
\end{equation}
and the differential equation on the maximal cut   \eqref{DE_dbox}
turns into the $\epsilon$-form:
\begin{equation}
  \label{eq:31}
\frac{\partial}{\partial \chi}  \left(
\begin{array}{c}
\tilde{J}[0,0] \\
\tilde{J}[1,0] \\
\end{array}
\right)
=
(D-4)\left(
\begin{array}{cc}
 \frac{1}{\chi  (\chi +1)} & \frac{1}{\chi +1} \\
 \frac{1}{2 \chi  (\chi +1)} & -\frac{1}{2 \chi  (\chi +1)} \\
\end{array}
\right)
\left(
\begin{array}{c}
\tilde{J}[0,0] \\
\tilde{J}[1,0] \\
\end{array}
\right) + \ldots \,.
\end{equation}

\subsection{Double box with one massive leg}
\begin{figure}[h]
  \centering
  \includegraphics{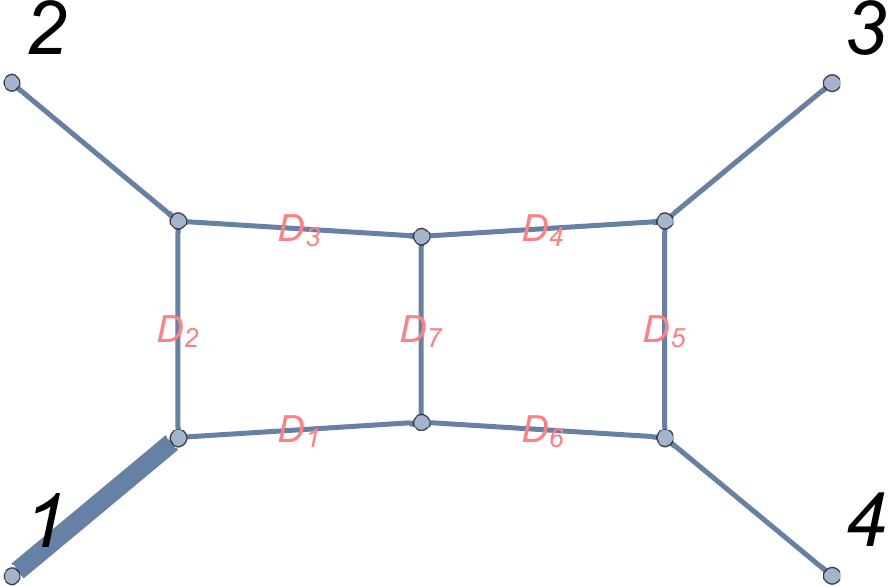}
\caption{Double box diagram with one external massive leg. This graph is produced by the package {\sc
  Azurite} .}
\label{dbox_1m_graph}
\end{figure}

We can use the same method to consider the maximal cut of the double box
with one massive external leg. The Feynman integral has the same inverse propagators
\eqref{dbox_prop}, and the kinematic conditions are $k_1^2=m_1^2$,
$k_2^2=k_3^2=k_4^2=0$. The two ISP are $D_8=(l_1+k_4)^2$, $D_9=(l_2+k_1)^2$ and the diagram is shown in figure \ref{dbox_1m_graph}.

Define $\chi=t/s$, $\kappa=m_1^2/s$, $x=z_8/s$ and $y=z_9/s$. The Baikov polynomial on the
maximal cut reads
\begin{equation}
  \label{eq:40}
  F=\frac{s^2 x y (\chi +\kappa  x-x y -x-y)}{4 \chi  (-\kappa +\chi +1)}\,.
\end{equation}
Consider the kinematic region $s>0$, $\chi>0$ and $\chi +1 -\kappa
>0$. Again, the integration region defined by $F\geq 0$ splits into four
subregions:
 \begin{itemize}
  \item[$\Omega_\text{I}$:] $x>0$, $y>0$, $\chi +\kappa  x-x y -x-y>0$,
  \item[$\Omega_\text{II}$:] $x>0$, $y<0$, $\chi +\kappa  x-x y -x-y<0$,
 \item[$\Omega_\text{III}$:] $x<0$, $y>0$, $\chi +\kappa  x-x y -x-y<0$,
  \item[$\Omega_\text{IV}$:] $x<0$, $y<0$, $\chi +\kappa  x-x y -x-y>0$.
 \end{itemize}
The subregions are shown in figure \ref{dbox_1m_region}. On the four subregions the conditions
$\mu_{11}(z)\geq 0$ and $\mu_{22}(z)\geq 0$ are satisfied. 
\begin{figure}[h!]
\centering
\includegraphics[scale=0.7]{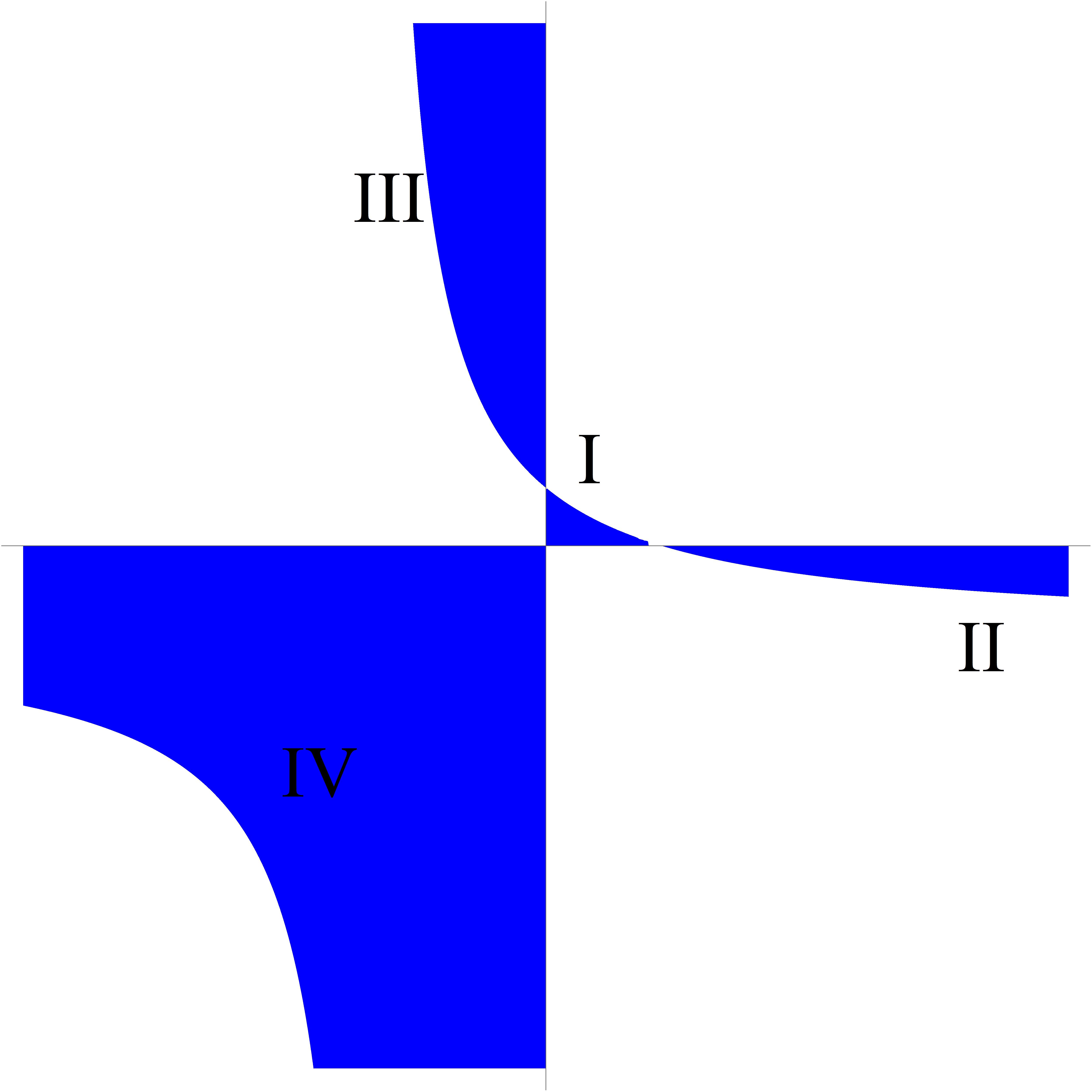}
\caption{Subregions for the integration of the double box
  diagram with one external massive leg on the maximal cut. The plot
  is for the kinematic configuration $s=1$, $\chi=1/3$ and
  $\kappa=1/2$. On the horizontal and vertical axis we have $x$ and $y$, respectively. Note that
  the subregions are not symmetric under a flip of the axes.}
\label{dbox_1m_region}
\end{figure}

The integrals under consideration are
\begin{equation}
  I[n_1,n_2,\ldots, n_7,-a,-b]\equiv \int \frac{d^D
    l_1}{\pi^{D/2}}\frac{d^D l_2}{\pi^{D/2}} \frac{D_8^a D_9^b}{D_1^{n_1}
    D_2^{n_2} \ldots D_7^{n_7}}\,.
\end{equation}
Again we define $J[a,b]\equiv I[1,1,1,1,1,1,1,-a,-b]$. The integration over $\Omega_\text{I}$ reads
\begin{align}
  \label{eq:22}
  J[a,b]\mc ^\text{(I)}&=\frac{ \Gamma
                         \left(\frac{D}{2}-2\right) \Gamma
                         \left(a+\frac{D}{2}-2\right) \Gamma
                         \left(b+\frac{D}{2}-2\right) s^{a+b-7}\chi ^{a+b+d-5}
                         \left(\frac{ 1 }{1-\kappa }\right)^{a+d-4}
                         }{16 \pi ^4  \chi ^5 \Gamma
                         (D-4)}\nonumber \\
& \times \, _2\tilde{F}_1\left(a+D-4,b+D-4;a+b+\frac{3
  D}{2}-6;\frac{\chi }{\kappa -1}\right)\,.
\end{align}
while the integration over $\Omega_\text{II}$ reads
\begin{align}
  \label{eq:41}
   J[a,b]\mc ^\text{(II)}&=\frac{(-1)^b(1-\kappa )^{-a+b}  \Gamma
     \left(\frac{D}{2}-2\right) \chi ^{a+\frac{D}{2}-3} \Gamma
     (-a-D+5) \Gamma \left(b+\frac{D}{2}-2\right) s^{a+b+D-7} }{16 \pi ^4 \Gamma
     (D-4)}\nonumber\\
&\times (-\kappa +\chi
     +1)^{2-\frac{D}{2}} \, _2\tilde{F}_1\left(-a-D+5,b+\frac{D}{2}-2;-a+b+1;\frac{\kappa -1}{\chi }\right)\,.
\end{align}
Similar to the massless case, the integrations over $\Omega_{\text{III}}$ and
$\Omega_{\text{IV}}$ are related to the previous two integrals through
\begin{align}
  \label{eq:42}
  J[a,b]\mc ^\text{(III)}&= J[a,b]\mc ^\text{(II)} \,,\\
J[a,b]\mc ^\text{(IV)}&= J[a,b]\mc ^\text{(I)}-2 \cos(\pi D)  J[a,b]\mc ^\text{(II)}\,.
\end{align}
Hence we can define the
$D$-dimensional maximal
cut as a list of two functions, $J[a,b]\mc \equiv \big( J[a,b]\mc
^\text{(I)},\  J[a,b]\mc ^\text{(II)}\big)$. We can see that in the
limit $\kappa\to 0$, these integrals become the maximal cut of the
massless double box. 

From Gauss' contiguous relations, we see that there are two master
integrals on the maximal cut, namely $J[0,0]$ and $J[1,0]$. For $i=$I,
II, III and IV,  $J[a,b]\mc ^\text{(i)}$ satisfy the same form of
function relations, for example, 
\begin{align}
  \label{eq:43}
  J[0,1]\mc ^\text{(i)}&= (1-\kappa ) J[1,0]\mc ^\text{(i)}\,,\nonumber \\
J[2,0]\mc ^\text{(i)}&=-\frac{(D-4) s^2 \chi  }{2 (D-3)
        (\kappa -1)}  J[0,0]\mc ^\text{(i)}-\frac{s (3 D \kappa -3
                       D-12 \kappa +2 \chi +12)}{2 (D-3) (\kappa -1)}
                       J[1,0]\mc ^\text{(i)}\,,\nonumber\\
J[2,1]\mc ^\text{(i)}&=-\frac{(3 D-10) s^3 \chi }{4
  (D-3)} J[0,0]\mc ^\text{(i)}-\frac{s^2 (9 D \kappa -2 d \chi -9 D-30
                       \kappa +8 \chi +30)}{4 (D-3)}J[1,0]\mc
                       ^\text{(i)}\,.
\end{align}
Again Gauss' contiguous relations provide IBPs on the maximal cut. We
can also check that these maximal cut functions satisfy dimension
shift identities. 

Define the $2\times2$ matrix
\begin{equation}
  \label{eq:29}
  S= \left( 
    \begin{array}{c}
      J[0,0]\mc\nonumber\\
J[1,0]\mc
    \end{array}
\right) \,.
\end{equation}
Explicitly, $S$ is the fundamental solution matrix of the differential
equations on the maximal cut,
\begin{equation}
\frac{\partial}{\partial \chi} \left(\begin{array}{c}
J[0,0]\\
J[1,0]
\end{array}\right)=
\left(
\begin{array}{cc}
 -\frac{\kappa  D-D-5 \kappa +\chi +5}{\chi  (-\kappa +\chi +1)} & \frac{(D-4) (\kappa -1)}{s (\kappa -\chi -1) \chi } \\
 -\frac{(D-4) s}{2 (\kappa -\chi -1)} & \frac{(D-4) (\kappa -1)}{2 \chi  (-\kappa +\chi +1)} \\
\end{array}
\right) \left(\begin{array}{c}
J[0,0]\\
J[1,0]
\end{array}\right) +\ldots\,,
\end{equation}
and
\begin{equation}
\frac{\partial}{\partial \kappa} \left(\begin{array}{c}
J[0,0]\\
J[1,0]
\end{array}\right)=
\left(
\begin{array}{cc}
 -\frac{D-4}{\kappa -\chi -1} & \frac{D-4}{s (-\kappa +\chi +1)} \\
 \frac{(D-4) s \chi }{2 (\kappa -1) (\kappa -\chi -1)} & \frac{\kappa  D-D-6 \kappa +2 \chi +6}{2 (\kappa -1) (\kappa -\chi -1)} \\
\end{array}
\right) \left(\begin{array}{c}
J[0,0]\\
J[1,0]
\end{array}\right) +\ldots\,.
\end{equation}

The leading coefficients of $S$ in the limit $D\to 4$ read
\begin{align}
  \label{eq:45}
  T=\left(
\begin{array}{cc}
 \frac{3}{4 \pi ^4 s^3 \chi } & \frac{1}{4 \pi ^4 s^3 \chi } \\
 -\frac{1}{4 \pi ^4 s^2 (\kappa -1)} & 0 \\
\end{array}
\right)\,.
\end{align}
Again, the transformation matrix is
\begin{align}
  \label{eq:46}
  \tilde T=T \left(\begin{array}{cc}
 0 & 1 \\
 1 & -3
\end{array}
\right) \cdot 4 \pi^4\,,
\end{align}
and the new master integrals are
\begin{equation}
  \label{eq:30}
  \tilde{J}[0,0] = s^3 \chi J[0,0]\,, \quad    \tilde{J}[1,0]= s^2 (1-\kappa)\chi J[1,0]\,.
\end{equation}
The differential equations turn into the $\epsilon$-form,
\begin{equation}
  \label{eq:31}
\frac{\partial}{\partial \chi}  \left(
\begin{array}{c}
\tilde{J}[0,0] \\
\tilde{J}[1,0] \\
\end{array}
\right)
=
(D-4)\left(
\begin{array}{cc}
 -\frac{ \kappa -1}{\chi  (-\kappa +\chi +1)} & -\frac{1}{\kappa -\chi -1} \\
 \frac{\kappa -1}{2 (\kappa -\chi -1) \chi } & -\frac{\kappa -1}{2 (\kappa -\chi -1) \chi } \\
\end{array}
\right)
\left(
\begin{array}{c}
\tilde{J}[0,0] \\
\tilde{J}[1,0] \\
\end{array}
\right) + \ldots \,,
\end{equation}
and
\begin{equation}
  \label{eq:31}
\frac{\partial}{\partial \chi}  \left(
\begin{array}{c}
\tilde{J}[0,0] \\
\tilde{J}[1,0] \\
\end{array}
\right)
=
(D-4)\left(
\begin{array}{cc}
 -\frac{1}{\kappa -\chi -1} & \frac{\chi }{(\kappa -1) (\kappa -\chi -1)} \\
 -\frac{1}{2 (\kappa -\chi -1)} & \frac{1}{2 (\kappa -\chi -1)} \\
\end{array}
\right)
\left(
\begin{array}{c}
\tilde{J}[0,0] \\
\tilde{J}[1,0] \\
\end{array}
\right) + \ldots \,.
\end{equation}

\subsection{Double box with two massive legs}
The maximal cut of the double box with two non-adjacent external  massive legs
($k_1^2\not=0$, $k_3^2\not=0$ or $k_1^2\not=0$, $k_4^2\not=0$)
again consists of hypergeometric $\,_2F_1$ functions, while the double box
with adjacent two external massive legs ($k_1^2\not=0$, $k_2^2\not=0$)
on the maximal cut yields Appell F1 functions. So in this
subsection, we focus on the latter case.  

\begin{figure}[h]
  \centering
  \includegraphics{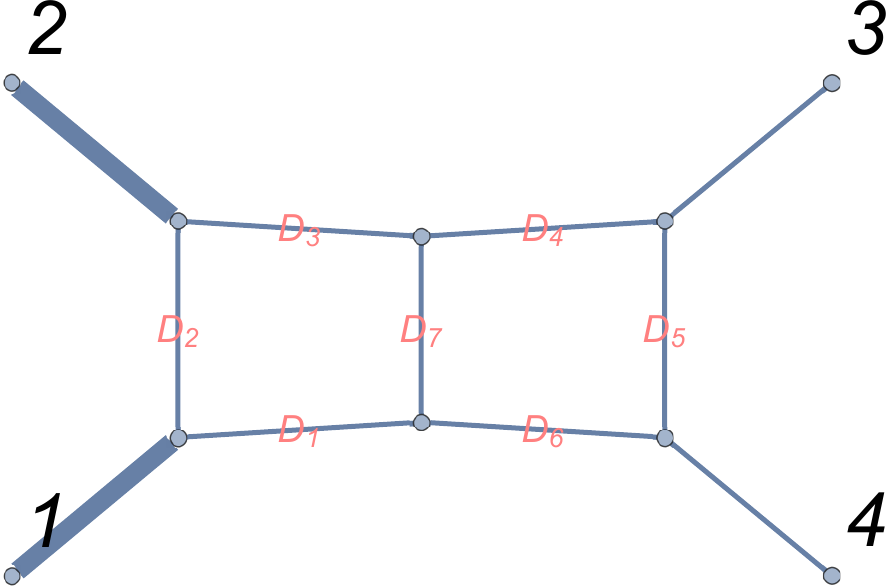}
\caption{Double box diagram with two adjacent external massive leg.  This graph is produced by the package {\sc
  Azurite}.}
\label{dbox_2m_graph}
\end{figure}

 The Feynman integral has the same inverse propagators
\eqref{dbox_prop}, and the kinematic conditions $k_1^2=m_1^2$,
$k_2^2=m_2^2$ and $k_3^2=k_4^2=0$. The two ISPs are $D_8=(l_1+k_4)^2$, $D_9=(l_2+k_1)^2$ and the diagram is shown in figure \ref{dbox_2m_graph}.
Define $\chi=t/s$, $\kappa_1=m_1^2/s$, $\kappa_2=m_2^2/s$, $x=z_8/s$
and $y=z_9/s$.  The Baikov polynomial on the maximal cut is
\begin{equation}
  \label{eq:52}
  F=-\frac{s^2 x \left(\kappa _1 \kappa _2 x+x y^2-\kappa _1 x y-\kappa _2 x y+x y+y^2-\chi  y\right)}{4 \left(-\kappa _1 \chi -\kappa _2 \chi +\kappa _1 \kappa _2+\chi ^2+\chi \right)}\,.
\end{equation}
For example, we consider the kinematic regime where $s>0$, $\kappa_1$ and
$\kappa_2$ are small. Specifically, the latter condition means,
\begin{align}
   -\kappa _1 \chi -\kappa _2 \chi +\kappa _1 \kappa _2+\chi ^2+\chi
  &>0 \label{dbox_2m_condition_1} ,\\
\kappa _1^2-2 \kappa _2 \kappa _1-2 \kappa _1+\kappa _2^2-2 \kappa
_2+1 &>0 \label{dbox_2m_condition_2}\,,
\end{align}
\eqref{dbox_2m_condition_1} is for the sign of the kinematic factor in
$F$. Note that on the curve $F(x,y)=0$, if $y\to c_1$ or $y\to c_2$,
\begin{align}
  \label{eq:48}
 c_1&=\frac{1}{2} \left(-1+\kappa _1+\kappa _2-\sqrt{\kappa _1^2-2 \kappa
  _2 \kappa _1-2 \kappa _1+\kappa _2^2-2 \kappa
  _2+1}\right) \,,\\
 c_2&=\frac{1}{2} \left(-1+\kappa _1+\kappa _2+\sqrt{\kappa _1^2-2 \kappa _2\kappa _1-2 \kappa _1+\kappa _2^2-2 \kappa _2+1}\right)\,,
\end{align}
then $x\to \infty$. It is then important to specify the sign of the
expression inside square roots, so we pick up the condition \eqref{dbox_2m_condition_2}. 

In this kinematic regime, the integration region defined by $F\geq 0$ splits into four
subregions:
 \begin{itemize}
  \item[$\Omega_\text{I}$:] $x>0$, $y>0$, $\kappa _1 \kappa _2 x+x y^2-\kappa _1 x y-\kappa _2 x y+x y+y^2-\chi  y<0$,
  \item[$\Omega_\text{II}$:] $x>0$, $y<0$, $\kappa _1 \kappa _2 x+x y^2-\kappa _1 x y-\kappa _2 x y+x y+y^2-\chi  y<0$,
 \item[$\Omega_\text{III}$:] $x<0$, $y>0$, $\kappa _1 \kappa _2 x+x y^2-\kappa _1 x y-\kappa _2 x y+x y+y^2-\chi  y>0$,
  \item[$\Omega_\text{IV}$:] $x<0$, $y<0$, $\kappa _1 \kappa _2 x+x y^2-\kappa _1 x y-\kappa _2 x y+x y+y^2-\chi  y>0$.
 \end{itemize}
The subregions are shown in figure \ref{dbox_1m_region}.
\begin{figure}[h!]
\centering
\includegraphics[scale=0.7]{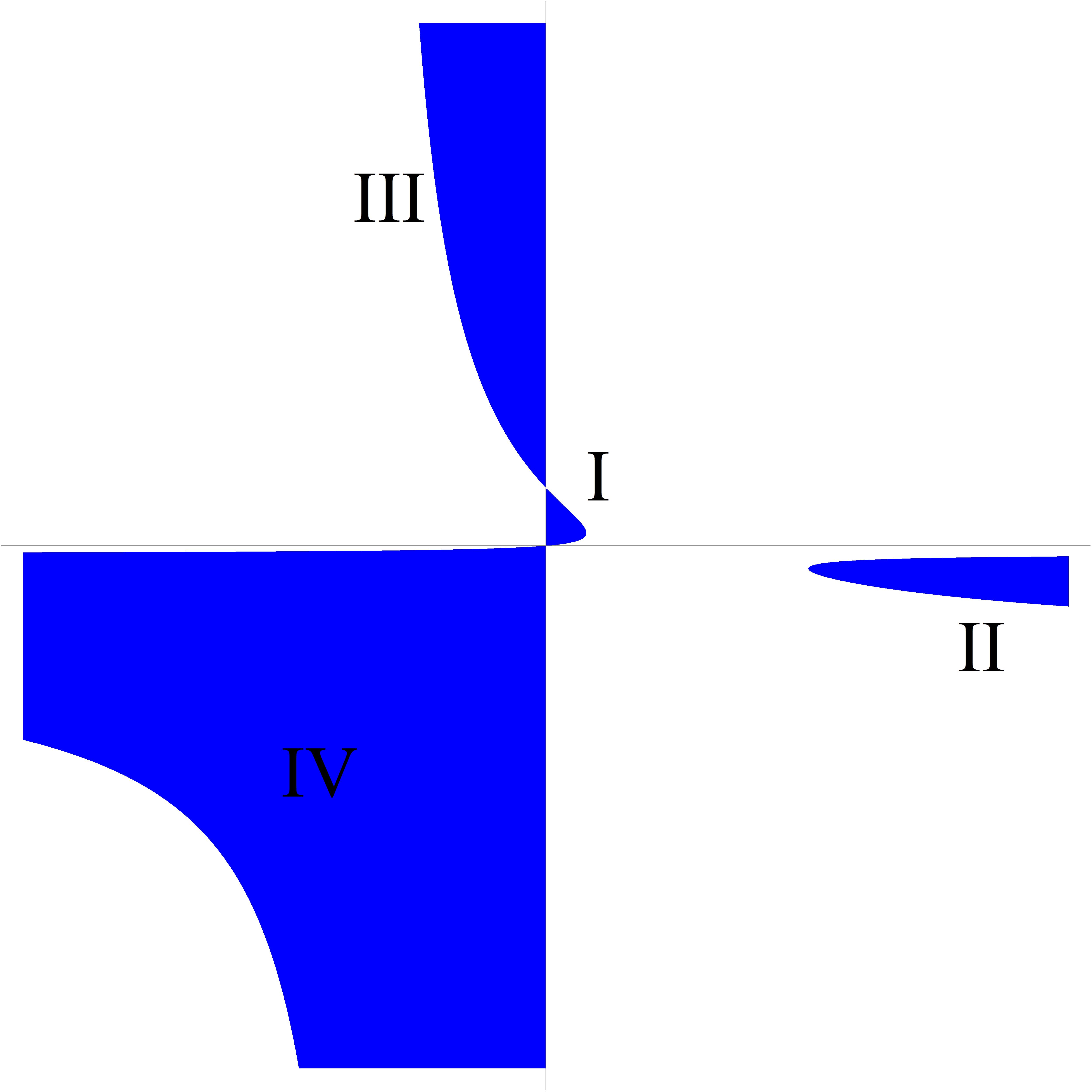}
\caption{Subregions for the integration of the double box
  diagram with two adjacent external massive leg on the maximal
  cut. The parameters for this plot are set as $s=1$, $\chi=1/3$,
  $\kappa_1=1/5$ and $\kappa_2=1/7$.
  On the horizontal and vertical axis we have $x$ and $y$, respectively.}
\label{dbox_1m_region}
\end{figure}
Explicitly, we can check that on the four subregions the conditions
$\mu_{11}(z)\geq 0$ and $\mu_{22}(z)\geq 0$ are satisfied. 

Again, the integrals under consideration are
\begin{equation}
  I[n_1,n_2,\ldots, n_7,-a,-b]\equiv \int \frac{d^D
    l_1}{\pi^{D/2}}\frac{d^D l_2}{\pi^{D/2}} \frac{D_8^a D_9^b}{D_1^{n_1}
    D_2^{n_2} \ldots D_7^{n_7}}\,.
\end{equation}
Again we define $J[a,b]\equiv I[1,1,1,1,1,1,1,-a,-b]$. The integration
over $\Omega_\text{I}$ contains a $y$-integral for
$y^{(\ldots)}(\chi-y)^{(\ldots)}(y-c_1)^{(\ldots)}(y-c_2)^{(\ldots)}$,
hence the result contains the Appell F1 function. Explicitly,
\begin{gather}
  \label{eq:53}
 J[a,b]\mc^{(\text{I})}=\left(\kappa _1 \kappa
      _2\right){}^{-a-\frac{D}{2}+2} \left(\left(\kappa _1-\chi \right)
      \left(\kappa _2-\chi \right)+\chi \right){}^{2-\frac{D}{2}}
    s^{a+b+D-7} \chi ^{2 a+b+2 D-9}\, \nonumber \\
  \times \frac{\Gamma \left(\frac{D}{2}-2\right) \Gamma
    \left(a+\frac{D}{2}-2\right) 
   \Gamma (a+b+D-4) }{16 \pi ^4
    \Gamma (D-4) \Gamma (2 a+b+2 D-8)}\,\nonumber \\
\times
F_1\left(a+b+D-4;\frac{1}{2} (2 a+D-4),\frac{1}{2} (2 a+D-4);2 a+b+2 D-8;w_1,w_2\right)\,.
\end{gather}
where $F_1$ is the Appell F1 function. The arguments $w_1$ and $w_2$
are defined as
\begin{equation}
  \label{eq:54}
  w_1=\frac{\chi}{\kappa_1 \kappa_2} c_1,\quad w_2=\frac{\chi}{\kappa_1 \kappa_2} c_2\,.
\end{equation}

The integration over $\Omega_\text{II}$ reads
\begin{gather}
  \label{eq:55}
  J[a,b]\mc^{(\text{II})}=
4^{a-2} (-1)^b \left(w_1-w_2\right){}^5 \left(1-w_2\right){}^{a+D-5} \left(\frac{w_1}{w_2}+\frac{w_2}{w_1}-2\right){}^{-a-\frac{D}{2}}  w_1^{-3} \left(-w_2\right){}^{-2 a-b-2 D+7}\nonumber \\
\times  \left(\left(\kappa _1-\chi \right)\left(\kappa _2-\chi \right)+\chi \right){}^{2-\frac{D}{2}}  (\kappa_1
  \kappa_2)^{-a-\frac{D}{2}+2}  s^{a+b+D-7} \chi ^{2 a+b+2 D-9} \frac{ \Gamma (-a-D+5)
 }{\pi ^3 \Gamma
  \left(\frac{D-3}{2}\right) \Gamma
  \left(-a-\frac{D}{2}+\frac{7}{2}\right)} \nonumber \\
\times F_1\left(-a-\frac{D}{2}+3;-a-b-D+5,-a-D+5;-2 a-D+6;1-\frac{w_2}{w_1},\frac{w_1-w_2}{w_1-w_1 w_2}\right)\,.
\end{gather}

The integration over $\Omega_\text{III}$ is
\begin{gather}
  \label{eq:56}
   J[a,b]\mc^{(\text{III})}=(-1)^{a} \left(\left(\kappa _1-\chi \right)
     \left(\kappa _2-\chi \right)+\chi \right){}^{2-\frac{D}{2}}
   s^{a+b+D-7} \chi ^{2 a+b+2 D-9} \left(\kappa _1 \kappa _2 w_1
     w_2\right){}^{-a-\frac{D}{2}+2} \nonumber\\
\frac{\Gamma \left(\frac{D}{2}-2\right) \Gamma \left(a+\frac{D}{2}-2\right)
\Gamma (-b-D+5)
F_1\left(-b-D+5;a+\frac{D}{2}-2,a+\frac{D}{2}-2;a-b+1;\frac{1}{w_1},\frac{1}{w_2}\right)}{16
\pi ^4 \Gamma (D-4) \Gamma (a-b+1)}\,.
\end{gather}

We can check that the last integration over $\Omega_\text{IV}$ is dependent, 
\begin{gather}
  \label{eq:57}
    J[a,b]\mc^{(\text{IV})} =  J[a,b]\mc^{(\text{I})} -2 \cos(\pi D)
    J[a,b]\mc^{(\text{III})} \,.
\end{gather}
Hence we define the maximal cut as the collection of three cut functions,
\begin{align}
  \label{eq:59}
  J[a,b]\mc=\bigg(J[a,b]\mc^{(\text{I})} , J[a,b]\mc^{(\text{II})} ,J[a,b]\mc^{(\text{III})}\bigg) \,.
\end{align}

There are three master integrals for this diagram on the maximal cut,
namely $J[0,0]$, $J[1,0]$ and $J[0,1]$. As was the case for the $\,_2F_1$ function,
the contiguous relations of Appell F1 functions generate IBP relations
on the maximal cut level. We see that the cut functions on
all four subregions,  $J[a,b]\mc^{(i)}$, $i=$I, II, III, IV, satisfy
IBP relations on the maximal cut. For example,
\begin{align}
  \label{eq:58}
  &J[1,1]\mc^{(i)}=\frac{1}{2} s^2 \chi  J[0,0]\mc^{(i)}+\frac{1}{2}
  \left(\kappa _1 +\kappa _2 -1\right)s J[1,0]\mc^{(i)}-s
                   J[0,1]\mc^{(i)}\,,\nonumber \\
&J[2,0]\mc^{(i)}=-\frac{(D-4) \left(\kappa _1+\kappa _2-1\right) s^2 \chi  }{2
  (D-3) \left(\kappa _1^2-2 \kappa _2 \kappa _1-2 \kappa _1+\kappa
  _2^2-2 \kappa _2+1\right)} J[0,0]\mc^{(i)} \,\nonumber \\
&-\frac{s}{2 (D-3) \left(\kappa
  _1^2-2 \kappa _2 \kappa _1-2 \kappa _1+\kappa _2^2-2 \kappa
  _2+1\right)} 
\bigg(2 D \kappa _1 \chi
  +2 D \kappa _2 \chi +D \kappa _1^2-2 D \kappa _1
-6 D \kappa _2
  \kappa _1+\nonumber\\
&D \kappa _2^2-2 D \kappa _2-2 D \chi +D-6 \kappa _1 \chi
  -6 \kappa _2 \chi -4 \kappa _1^2+20 \kappa _2 \kappa _1+8 \kappa
  _1-4 \kappa _2^2+8 \kappa _2+6 \chi -4\bigg)
J[1,0]\mc^{(i)}\,\nonumber\\
&+\frac{(D-4) s \left(\kappa _1+\kappa _2-\chi
  -1\right)}{(D-3) \left(\kappa _1^2-2 \kappa _2 \kappa _1-2 \kappa
  _1+\kappa _2^2-2 \kappa _2+1\right)} J[0,1]\mc^{(i)}\,.
\end{align}
Similarly, $J[a,b]\mc^{(i)}$, $i=$I, II, III, IV, satisfy
dimension-shift identities because of Appell F1 functions' contiguous relations.

\begin{figure}[h]
  \centering
  \includegraphics[scale=0.9]{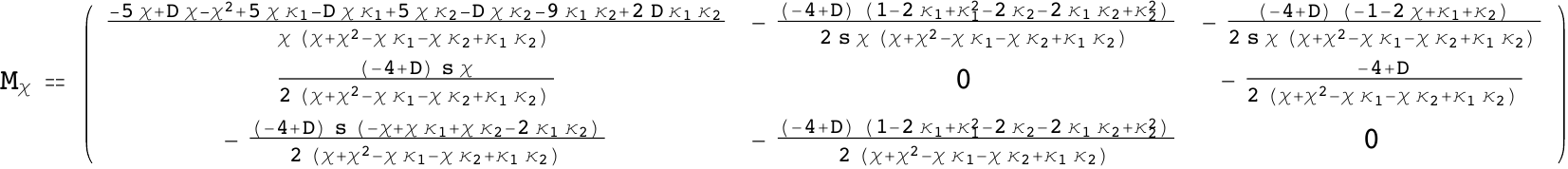}
\includegraphics[scale=0.62]{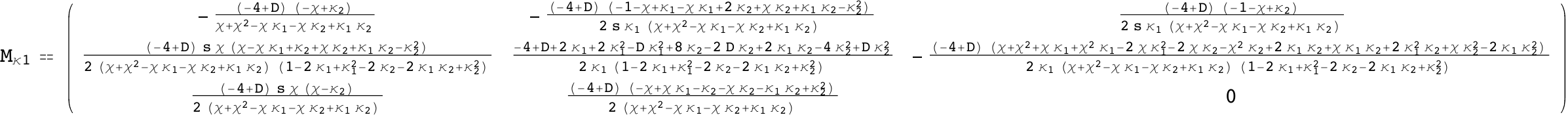} 
\includegraphics[scale=0.62]{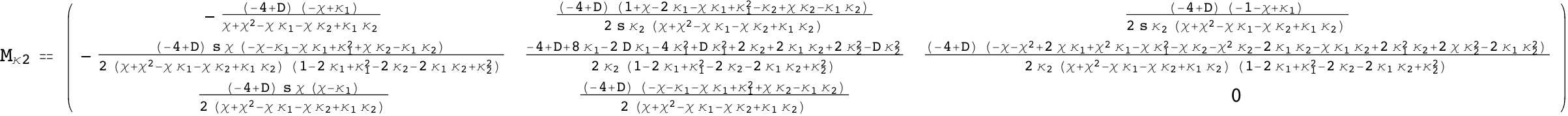}
  \caption{Matrices for differential equations on the maximal cut for the double box diagram with two adjacent
    massive legs.}
\label{dbox_2m_matricies}
\end{figure}

Let $I=(J[0,0],J[1,0],J[0,1])^T$. $I$ should satisfy the differential equations on
the maximal cut,
\begin{equation}
  \label{eq:60}
  \frac{\partial }{\partial \chi }I =M_\chi I + \ldots,\quad 
\frac{\partial }{\partial \kappa_1 }I =M_{\kappa_1} I + \ldots,\quad
\frac{\partial }{\partial \kappa_2 }I =M_{\kappa_2} I + \ldots \,,
\end{equation}
where the $3\times 3$ matrices are given in figure \ref{dbox_2m_matricies}.
Explicitly, for any $i=$I,
II, III, IV, the vector $(J[0,0]\mc^{(i)}, J[1,0]\mc^{(i)}, J[0,1]\mc^{(i)})^T$ solves the differential equations on
the maximal cut. Furthermore, we find that the $3\times 3$ matrix
\begin{equation}
  \label{eq:63}
  S=
\left(
  \begin{array}{c}
    J[0,0]\mc\\
J[1,0]\mc\\
J[0,1]\mc
  \end{array}
\right)\,,
\end{equation}
is the fundamental solution matrix for all these differential
equations and the Wronskian is nonzero.

\begin{figure}[h]
  \centering
  \includegraphics[scale=1]{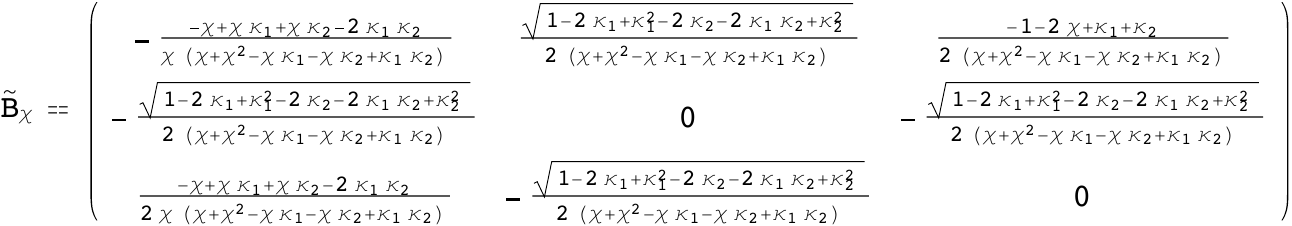}
\includegraphics[scale=0.62]{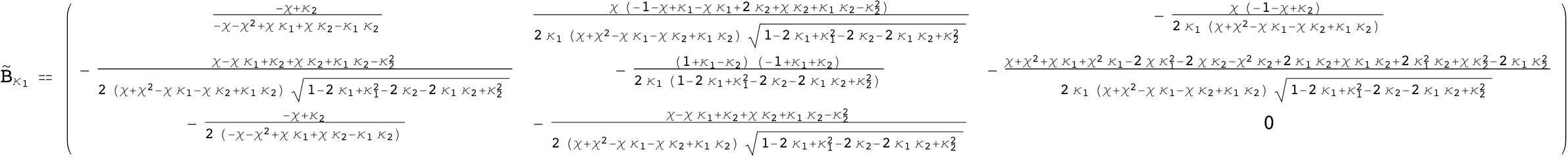} 
\includegraphics[scale=0.62]{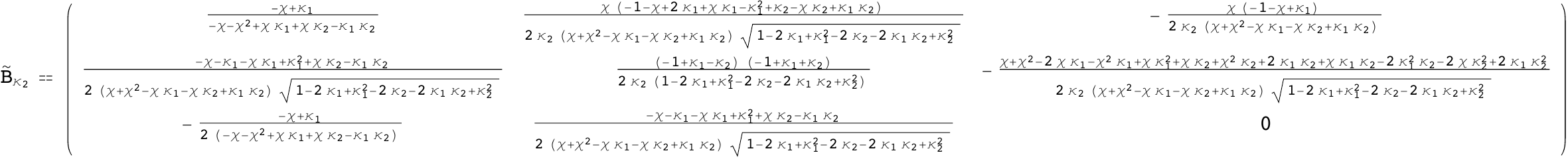}
  \caption{Matrices for the $\epsilon$-form differential equations on
    the maximal cut for the double box diagram with two adjacent
    massive legs.}
\label{dbox_2m_matricies_epsilon}
\end{figure}

The leading coefficients of $S$ in the limit $D\to 4$ are
\begin{equation}
 T=\left(
\begin{array}{ccc}
 \frac{1}{2 \pi ^4 s^3 \chi } & 0 & \frac{1}{4 \pi ^4 s^3 \chi } \\
 0 & -\frac{1}{4 \pi ^4 \sqrt{\kappa _1^2-2 \kappa _2 \kappa _1-2 \kappa _1+\kappa _2^2-2 \kappa _2+1} s^2} & 0 \\
 \frac{1}{4 \pi ^4 s^2} & 0 & \frac{1}{4 \pi ^4 s^2} \\
\end{array}
\right)\,.
\end{equation}
By a simple column operation, the transformation matrix can be chosen
as
\begin{equation}
  \tilde T=T.
\left( 
 \begin{array}{ccc}
    1 & 0 &1\\
    0& 1 & 0\\
    -1& 0 & -2
  \end{array}
\right) 4 \pi^4=
\left(
\begin{array}{ccc}
 \frac{1}{s^3 \chi } & 0 & 0 \\
 0 & -\frac{1}{\sqrt{\kappa _1^2-2 \kappa _2 \kappa _1-2 \kappa _1+\kappa _2^2-2 \kappa _2+1} s^2} & 0 \\
 0 & 0 & -\frac{1}{s^2} \\
\end{array}
\right)\,.
\end{equation}
So we can redefine the basis as
\begin{equation}
  \label{eq:44}
  \tilde I=(s^3 \chi J[0,0], -s^2\sqrt{\kappa _1^2-2 \kappa _2 \kappa _1-2
    \kappa _1+\kappa _2^2-2 \kappa _2+1} J[1,0],-s^2 J[0,1])^T\,.
\end{equation}
The differential equations on the maximal cut for $\tilde I$ is in the
$\epsilon$-form,
\begin{equation}
  \frac{\partial }{\partial \chi } \tilde I =(D-4)\tilde B_\chi \tilde I + \ldots,\quad 
\frac{\partial }{\partial \kappa_1 } \tilde  I =(D-4)\tilde B_{\kappa_1} \tilde I + \ldots,\quad
\frac{\partial }{\partial \kappa_2 }\tilde I =(D-4)\tilde B_{\kappa_2}\tilde I + \ldots \,,
\end{equation}
where the matrices $\tilde B_\chi$, $\tilde  B_{\kappa_1}$ and $\tilde B_{\kappa_2}$ are
given in figure \ref{dbox_2m_matricies_epsilon}. Again $\ldots$
stands for integrals with fewer propagators.

\section{Nonplanar maximal cuts with two ISPs}
We are also interested in confirming that nonplanar integrals are equally
amenable to the $D$-dimensional maximal cut procedure demonstrated for one-loop
and planar two-loop integrals throughout the previous sections. More
specifically, we analyze the maximal cut regions for the purely massless
nonplanar double box, evaluate the two-fold integral in the post-hepta-cut
degrees of freedom for each subregion separately, and finally investigate the
properties of the resulting maximal cut expression.

\subsection{Massless nonplanar double box}
\begin{figure}[!h]
\centering
\includegraphics[scale=0.7]{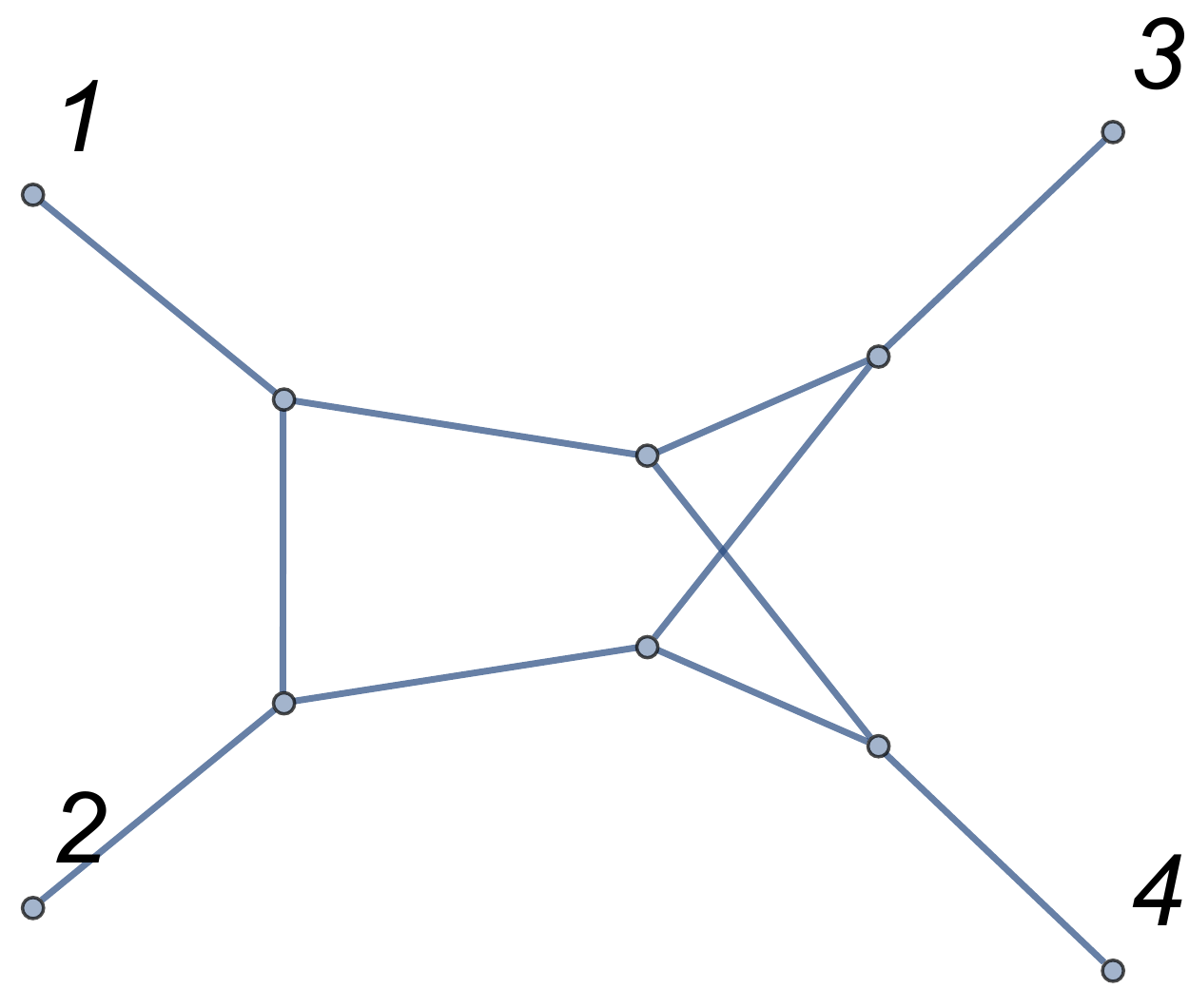}
\caption{The nonplanar double box with four external lines.}
\label{FIG_XBOX}
\end{figure}

We adopt the conventions for the external kinematics from the preceeding
section. Consider now a generic purely massless $D$-dimensional nonplanar
double-box integral,
\begin{align}
X[n_1,n_2,\dots,n_7,-a,-b] = 
\int\frac{d^Dl_1}{\pi^{D/2}}
\int\frac{d^Dl_2}{\pi^{D/2}}
\frac{D_8^aD_9^b}{D_1^{n_1}D_2^{n_2}\cdots D_7^{n_7}}\,,
\end{align}
specified by the seven inverse propagators,
\begin{gather}
D_1 = \ell_1^2\,, \quad
D_2 = (\ell_1-k_1)^2\,, \quad
D_3 = (\ell_1-k_1-k_2)^2\,, \quad
D_4 = \ell_2^2\,, \\
D_5 = (\ell_2-k_4)^2\,, \quad
D_6 = (\ell_1+\ell_2)^2\,, \quad
D_7 = (\ell_1+\ell_2+k_3)^2\,,
\end{gather}
and two ISPs, which may be picked up as the conventional propagator-like forms,
\begin{align}
D_8 = (\ell_1+k_1)^2\,, \quad
D_9 = (\ell_2+k_4)^2\;.
\end{align}
The momentum flow conventions corresponding to the integral under consideration
are shown in figure~\ref{FIG_XBOX}.

\begin{figure}[!h]
\centering
\includegraphics[scale=0.7]{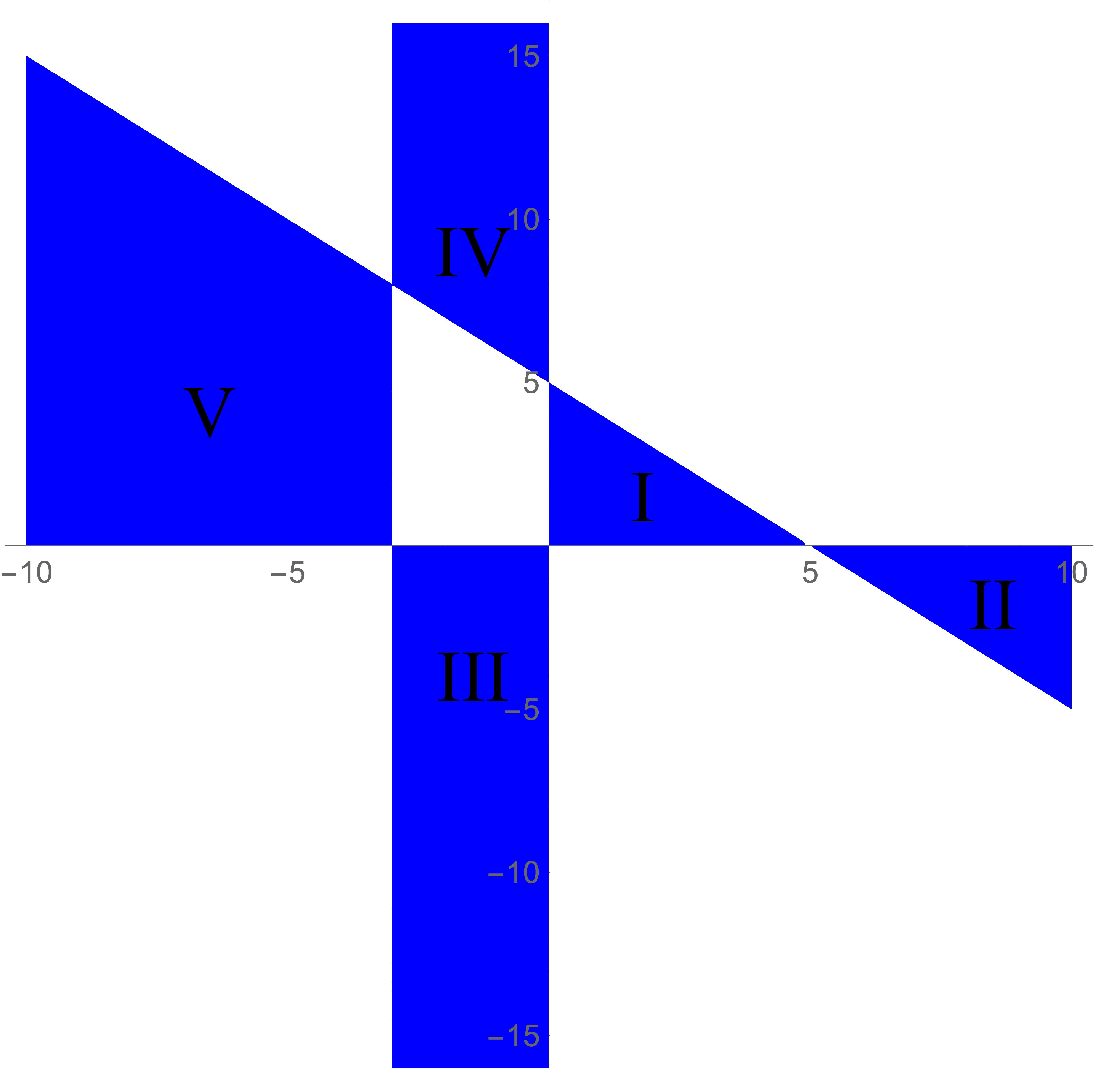}
\caption{The five regions $\Omega_I,\dots,\Omega_V$ for the maximally cut
nonplanar double box with massless internal and external kinematics. The colored
area corresponds to $F\geq 0$, where $F$ is the Baikov polynomial. The numeric
conditions for this plot are $s = 3$ and $t = 5$.}
\label{FIG_XBOX_REGIONS}
\end{figure}

The maximal cut evaluated in the Baikov representation reads
\begin{align}
	X[a,b]\mc^{(\Omega)} = \frac{2^{D-5}}{\pi^4\Gamma(D-4)\det G_3}
\int_\Omega dz_8dz_9 z_8^az_9^b F(z_8,z_9)^{\frac{D-6}{2}}
\,,
\label{EQ_XBOX_MAXCUT}
\end{align}
where 
\begin{align} 
F(z_8,z_9) = 
\frac{z_8 z_9\left(s+z_8\right)\left(s\chi-z_8-z_9\right)}
{4s^2\chi(\chi+1)}\;.
\end{align}
This expression for $F$ parallels eq.~\eqref{eq:20}. The domain of integration
$\Omega$ is again defined as the region of the $(z_8,z_9)$-plane where the cut
Baikov polynomial is nonnegative. As shown in figure~\ref{FIG_XBOX_REGIONS}, for
the problem at hand, $\Omega$ is divided into five simple subregions. These
subregions take the shape of triangles and rectangles, corresponding to the
inequalities:
\begin{itemize}
\item[$\Omega_\text{I}$:] $z_8>0\,,\;z_9>0\,,\;s+z_8>0\,,\;t-z_8-z_9>0\,,$
\item[$\Omega_\text{II}$:] $z_8>0\,,\;z_9<0\,,\;s+z_8>0\,,\;t-z_8-z_9<0\,,$
\item[$\Omega_\text{III}$:] $z_8<0\,,\;z_9<0\,,\;s+z_8>0\,,\;t-z_8-z_9>0\,,$
\item[$\Omega_\text{IV}$:] $z_8<0\,,\;z_9>0\,,\;s+z_8>0\,,\;t-z_8-z_9<0\,,$ 
\item[$\Omega_\text{V}$:] $z_8<0\,,\;z_9>0\,,\;s+z_8<0\,,\;t-z_8-z_9>0\,.$
\end{itemize}

We now calculate the double integral for each subregion separately. For example,
the integration over region $I$ is parametrized as follows,
\begin{align}
X[a,b]\mc^{(I)} = {} & 
\frac{2^{D-5}}{\pi^4\Gamma(D-4)\det G_3}
\int_0^tdz_8\int_0^{t-z_8}dz_9 
F(z_8,z_9)^{\frac{D-6}{2}}z_8^az_9^b
\;.
\end{align}
In this instance and also for the remaining four subregions the result is
straightforwardly obtained in the form of $_2F_1$ hypergeometric functions
multiplied by kinematic invariants and some Gamma functions as expected. The
result for region $I$ can be written as
\begin{align}
X[a,b]\mc^{(I)} = {} &
\frac{2^{D-5}}{\pi^4\Gamma(D-4)}
s^{a+b+D-7}\chi^{a+b+D-5}(1+\chi)^{2-\frac{D}{2}}
\Gamma\Big(a+\frac{D}{2}-2\Big)
\Gamma\Big(b+\frac{D}{2}-2\Big)
\nn \\[1mm] & \qquad\times
\Gamma\Big(\frac{D}{2}-2\Big)
\,_2\tilde{F}_1
\Big(3-\frac{D}{2},a+\frac{D}{2}-2;a+b+\frac{3D}{2}-6;-\chi\Big)
\label{EQ_XBOX_R1}
\,.
\end{align}
Completely analogously, for region $II$,
\begin{align}
\label{EQ_XBOX_R2}
X[a,b]\mc^{(II)} = {} & 
\frac{2^{D-5}}{\pi^4\Gamma(D-4)}
(-1)^{b+D}s^{a+b+D-7}
\chi^{a+b+\frac{3D}{2}-8}
(1+\chi)^{2-\frac{D}{2}}
\Gamma(10-a-b-2D)
\nn \\[1mm] & \;\;\times
\Gamma\Big(\frac{D}{2}-2\Big)
\Gamma\Big(b+\frac{D}{2}-2\Big)
\,_2\tilde{F}_1
\Big(10-a-b-2D,3-\frac{D}{2};6-a-D;-\frac{1}{\chi}\Big)
\,.
\end{align}
Next, we present the results for regions $III$ and $IV$,
\begin{align}
\label{EQ_XBOX_R3}
&X[a,b]\mc^{(III)} =
\frac{2^{5-D}}{\pi^4\Gamma(D-4)}
(-1)^{1+a+b}s^{a+b+D-7}
\chi^{b+\frac{D}{2}-3}
(1+\chi)^{2-\frac{D}{2}} 
\Gamma(5-b-D)
\\ & \quad \times
\frac{\Gamma\Big(a+\frac{D}{2}-2\Big)
\Gamma\Big(b+\frac{D}{2}-2\Big)
\Gamma\Big(\frac{D}{2}-2\Big)}{
\Gamma\Big(3-\frac{D}{2}\Big)}
\,_2\tilde{F}_1\Big(5-b-D,a+\frac{D}{2}-2;a+D-4;-\frac{1}{\chi}\Big)\,,
\nn
\end{align}
\begin{align}
\label{EQ_XBOX_R4}
&X[a,b]\mc^{(IV)} = 
\frac{2^{D-5}}{\pi^4\Gamma(D-4)}
(-1)^{a+1}s^{a+b+D-7}
\chi^{b+\frac{D}{2}-3}
(1+\chi)^{2-\frac{D}{2}}
\Gamma(5-b-D)
\nn \\ &\qquad\quad\times
\frac{\Gamma\Big(a+\frac{D}{2}-2\Big)
\Gamma^2\Big(\frac{D}{2}-2\Big)}{
\Gamma\Big(3-b-\frac{D}{2}\Big)}
\,_2\tilde{F}_1\Big(5-b-D,a+\frac{D}{2}-2;a+D-4;-\frac{1}{\chi}\Big)\,.
\end{align}
Finally, the integration over region $V$ yields a slightly more complicated
expression involving multiple hypergeometric functions. As we shall see below,
the fully simplified output from Mathematica can be further manually reduced to
a very compact form.

By inspection of eqs.~\eqref{EQ_XBOX_R1}-\eqref{EQ_XBOX_R4} along with the
expression obtained for region $V$, we see that the maximal cut of the nonplanar
double box basically gives rise to Gauss' $_2F_1$ hypergeometric functions with
four distinct quadruples of indices, namely,
\begin{gather}
_2\tilde{F}_1\Big(3-\frac{D}{2},a+\frac{D}{2}-2;a+b+\frac{3D}{2}-6;-\chi\Big)\,,
\\
_2\tilde{F}_1\Big(10-a-b-2D,3-\frac{D}{2};6-a-D;-\frac{1}{\chi}\Big)\,, \\
_2\tilde{F}_1\Big(5-b-D,a+\frac{D}{2}-2;a+D-4;-\frac{1}{\chi}\Big)\,, \\
_2\tilde{F}_1\Big(10-a-b-2D,5-b-D;8-a-b-\frac{3D}{2};-\chi\Big)\;.
\end{gather}
Again, it is however trivial to reexpress two of the hypergeometric functions in
terms of the remaining two. For instance, we can choose the integration over
regions $I$ and $II$ as the principal results, and simply define the maximal cut
function as
\begin{align}
X[a,b]\mc\equiv\Big(X[a,b]\mc^{(I)},X[a,b]\mc^{(II)}\Big)\;.
\label{EQ_XBOX_MAXCUT}
\end{align}
Given that $a$ and $b$ always assume integer values we found the following
remarkable simplifications for regions $III$, $IV$ and $V$,
\begin{align}
X[a,b]\mc^{(III)} = {} &
e^{i\pi D}
X[a,b]\mc^{(I)}+
2\cos \pi D
X[a,b]\mc^{(II)} = 
X[a,b]\mc^{(IV)}\,, \\
%%%
X[a,b]\mc^{(V)} = {} &
e^{i\pi D}
X[a,b]\mc^{(I)}+
(1+2\cos \pi D)
X[a,b]\mc^{(II)}\;.
\end{align}
It is readily verified that all the displayed coefficients can be regarded as
merely constants in connection with integral relations, as they are independent
of $a$ and $b$, and invariant under dimensional shifts $D\to D\pm 2$.

As for the planar double box, Gauss' contiguous relations immediately provide
all necessary information about the IBPs on the maximal cut by the same
argument. From the simple structure of the maximal cut we know that
$X[a,b]\mc^{(i)}$ can be reduced to a linear combination of, say,
$X[0,0]\mc^{(i)}$ and $X[1,0]\mc^{(i)}$, corresponding to the usual scalar and
rank-1 tensor master integrals. This observation holds for any of the five
subregions individually. For example, up to rank-2 tensors, for $i =
I,II,III,IV,V$,
\begin{align}
X[0,1]\mc^{(i)} = {} &
+\frac{t}{2}X[0,0]\mc^{(i)}-
\frac{1}{2}X[1,0]\mc^{(i)}\,, \\
%%%
X[1,1]\mc^{(i)} = {} &
-\frac{1}{8}s\chi X[0,0]\mc^{(i)}+
\frac{1}{8}s(2\chi+3)X[1,0]\mc^{(i)}\,, \\
%%%
X[2,0]\mc^{(i)} = {} &
+\frac{1}{4}s^2\chi X[0,0]\mc^{(i)}-
\frac{1}{4}s(3-2\chi)X[1,0]\mc^{(i)}\,, \\
%%%%
X[0,2]\mc^{(i)} = {} &
+\frac{(D-2)s^2\chi(4\chi+1)}{16(D-3)}X[0,0]\mc^{(i)}-
\frac{3(D-2)s(2\chi+1)}{16(D-3)}X[1,0]\mc^{(i)}\;.
\end{align}
An elementary manipulation of eq.~\eqref{EQ_XBOX_MAXCUT} enables us to also
include nonplanar double box integrals with doubled (or simply arbitrary powers
of) propagators in the present setup. Here, however, we refrain for brevity from
giving any examples along this direction. Instead, we verify that our maximal
cut inherits the dimension shifting properties satisfied by the uncut integral.
Explicitly, it can be shown that the maximal cut satisfies the raising 
recurrence relation,
\begin{align}
&X[0,0]\mc^{(i)}(D-2) = \nn \\ & 
\frac{D-5}{(D-6) s^2 \chi ^2 (\chi +1)^2}\Big(
2 (3 D-16) (10 D-47) \chi +3 (9 (D-10) D+224) \nn \\
& \qquad\qquad
-16 (D-5)^2 \chi ^4-8 (D-5) (2 D-11) \chi ^3+24 (D-5)^2 \chi ^2\Big)
X[0,0]\mc^{(i)}(D)\nn \\[1mm] &\quad
+\frac{2 (D-5) (2 D-9) (2 \chi +1) (D (4 \chi  (\chi +1)+9)-4 (5 \chi  (\chi
+1)+12))}{(D-6) s^3 \chi ^2 (\chi +1)^2}
X[1,0]\mc^{(i)}(D)\,,
\end{align}
and also the lowering recurrence relation,
\begin{align}
&X[0,0]\mc^{(i)}(D+2) = \nn \\
&\quad
\frac{(D-4) s^2 (D (4 (\chi -2) \chi -3)+4 \chi  (7-3 \chi )+10)}{128 (D-3)^3 (2
D-7) (\chi +1)}
X[0,0]\mc^{(i)}(D)
\nn \\ & \qquad +
\frac{(D-4) s (2 \chi +1) (D (4 \chi  (\chi +1)+9)-6 (2 \chi  (\chi +1)+5))}{128
(D-3)^3 (2 D-7) \chi  (\chi +1)}
X[1,0]\mc^{(i)}(D)\;.
\end{align}
These equations are true for any of the five subregions. Similar identities hold
for all other nonplanar double box integrals. All integral relations inferred
from the maximal cut are validated by FIRE \cite{Smirnov:2005ky,Smirnov:2006tz, Smirnov:2008iw, Smirnov:2014hma} and Azurite \cite{Georgoudis:2016wff}.

Let us finally discuss differential equations obeyed by the nonplanar double
box integrals and their relation to the maximal cut. From the IBP relations it 
has already been established that
\begin{equation}
  \label{DE_xbox}
  \frac{d}{d\chi} \left(\begin{array}{c}
\tilde J[0,0]\\
\tilde J[1,0]
\end{array}\right) =
\left(
\begin{array}{cc}
 \frac{D-\chi -5}{\chi  (\chi +1)} & \frac{2 D-9}{s \chi  (\chi +1)} \\
 \frac{(D-4) s}{2 (\chi +1)} & \frac{2 \chi  D-D-10 \chi +4}{2 \chi  (\chi +1)} \\
\end{array}
\right)
\left(\begin{array}{c}
\tilde J[0,0]\\
\tilde J[1,0]
\end{array}\right) +\cdots\,,
\end{equation}
suppressing integrals with fewer than seven propagators. We have explicitly
checked that our maximal cut indeed solves this differential equation, region by
region. Moreover, the maximal cut functions \eqref{EQ_XBOX_MAXCUT} again form
the Wronskian matrix $S$ associated with this system of differential equations.
Defining $S$ as follows,
\begin{equation}
  S= \left( 
    \begin{array}{c}
      X[0,0]\mc\nonumber\\[1mm]
      X[1,0]\mc
    \end{array}
\right)\,,
\end{equation}
the Wronskian determinant is found to be nonvanishing,
\begin{align}
\det S\propto \chi ^{\frac{D}{2}-3} (\chi +1)^{\frac{D}{2}-3}\;.
\end{align} 
This ensures that $X[a,b]\mc^{(I)}$ and $X[a,b]\mc^{(II)}$ are linearly
independent functions of $\chi$ as previously anticipated, and furthermore
confirms that the columns of $S$ form the two fundamental solutions to
eq.~\eqref{DE_xbox}. The leading terms of $S$ in the limit $D\to 4$,
\begin{align}
T = 
\frac{16}{\pi^4}
\left(
\begin{array}{cc}
 \frac{2 \chi +3}{s^3 \chi  (\chi +1)} & -\frac{1}{s^3 \chi  (\chi +1)} \\
 \frac{1}{s^2 (\chi +1)} & -\frac{1}{s^2 (\chi +1)} \\
\end{array}
\right)
\end{align}
again help us to identify a new set of master integrals in order to transform
eq.~\eqref{DE_xbox} to $\epsilon$-form. We may include a trivial redefinition
of $T$ and take the transformation matrix as
\begin{align}
T' = 
T
\left(
\begin{array}{cc}
 1 & -4 \\
 0 & -4 \\
\end{array}
\right)
\cdot\frac{\pi^4}{16}
\,,
\end{align}
which by eq.~\eqref{eq:39} implies that
\begin{align}
\frac{d}{d\chi} \left(\begin{array}{c}
\tilde J_1 \\
\tilde J_2
\end{array}\right) =
(D-4)
\left(
\begin{array}{cc}
 \frac{2 \chi +1}{\chi  (\chi +1)} & -\frac{4}{\chi } \\
 \frac{1}{2 (\chi +1)} & -\frac{2 \chi +1}{2 \chi  (\chi +1)} \\
\end{array}
\right)
\left(\begin{array}{c}
\tilde J_1 \\
\tilde J_2
\end{array}\right)\,,
\end{align}
where the modified masters $\tilde J_1$ and $\tilde J_2$ denote the following
mixture of the FIRE basis integrals,
\begin{align}
\tilde J_1\equiv s^2 (1+\chi)J[1,0]\,, \quad 
\tilde J_2\equiv -\frac{1}{8}s^3\chi J[0,0]+\frac{1}{8}s^2(2\chi+3)J[1,0] = 
s J[1,1]\;.
\end{align}
The new differential equation is obviously in $\epsilon$-form for $D = 4-2\epsilon$.

In summary, we have explicitly verified that all salient features of the maximal
cut of the planar double box carry over immediately to the nonplanar double box.
Our example here only covered the purely massless case, but we also have
succesfully checked several configurations with massive external momenta.

%%%%%%%%%%%%%%%%%%%%%
\section{Conclusion}
%%%%%%%%%%%%%%%%%%%%%%%
In this paper we have presented a precise and consistent technique for
evaluating generalized cuts of multiloop Feynman integrals, properly
regularized in $D$ dimensions. Our method relies on the Baikov representation,
which makes the effect of taking these generalized cuts in arbitrary dimension
manifest. We have given examples of the method for several integral topologies
with various kinematic configurations, including the one-loop box, two-loop
sunset, planar double box and nonplanar double box.

The simplest instance is the maximal cut of a $k$-propagator integral realized
by the multivariate residue of the integrand at $z_1 = \cdots = z_k = 0$, the
$z_i$'s being the Baikov variables. In general the maximal cut leaves a
multi-fold integral over a domain $\Omega$ defined by the intersection of the cut
hyperplanes and the region $A$ where the Baikov kernel $F$ is nonnegative.
The remaining integration may be carried out over any subregion of $\Omega$ with 
$F = 0$ on the boundary. We refer to the result as the maximal cut function on
subregion $j$; this can be viewed as a generalization of the notion of a
composite leading singularity. 

The maximal cut function satisfies the same form of integral relations, such
as IBPs and dimension shift identities, and differential equations as the uncut
integral, region by region. In our examples, the maximal cut functions are
compact analytic expressions involving Gamma functions and (generalized-)
hypergeometric functions. The integral identities on the maximal cut hence
immediately correspond to relations among these special functions, namely recurrence
relations and Gauss' contiguous relations.

One of the principal features of all our examples is that an integral topology
with $m$ master integrals has precisely $m$ linearly independent maximal cut
function series. For instance, the purely nonplanar double box, with
real kinematics, gives rise to five
subregions, but only two linearly independent maximal cut
functions. This number
agrees with the number of master integrals. We have explicitly shown that
the linearly independent maximal cut functions form the Wronskian matrix $S$
for the system of differential equations on the maximal cut. Moreover, we have
in detail demonstrated that the leading terms of $S$ provide a transformation
matrix for differential equation into to canonical (epsilon) form near four
dimensions.

From the viewpoint of the differential equation of Feynman
integrals without cut, these maximal-cut
functions form the fundamental solution set of the ``homogenous'' equation. To solve the complete
differential equation system, can be understood as
solving for an ``inhomogenous'' differential equation. Hence it is
expected that these
functions would appear in the complete expression of Feynman
integrals. 

This work brings inspiration for further advances along the direction of
multiloop generalized unitarity with arbitrary spacetime dimension.
\begin{itemize}
\item In this paper, we simply consider the integration regions
  corresponding to the
  real loop momenta and find that it is enough to get the complete
  solutions for the differential equations on the maximal cut. For
  more complicated integrals, we may also consider complex loop
  momenta and integration regions for complex Baikov variables. (For a
  region to be valid, we still require that on the boundary the
  Baikov polynomial vanishes, i.e., $F=0$.)

\item It would be very interesting to study the maximal cut of elliptic
  Feynman integrals \cite{Laporta:2004rb, Bloch:2013tra,
    Adams:2013kgc, Adams:2014vja, Adams:2015gva, Bloch:2016izu, Passarino:2016zcd,
    Kalmykov:2016lxx, vonManteuffel:2017hms, Adams:2017tga} with
  arbitrary spacetime dimension. The $4D$ maximal cut of an elliptic
  double box was studied via Weierstrass elliptic functions
  \cite{Sogaard:2014jla}. The $(D-2)$-iterative form of elliptic differential
  equations were studied in ref.~\cite{Adams:2015ydq,
    Tancredi:2015bvi, Remiddi:2016gno}. We expect that our method
  combined with
  integrals over the fundamental domain of elliptic curves, would
  provide the maximal cut of elliptic Feynman integrals in a closed
  form of $D$.

\item It is also interesting to see the non-maximal cut in
  $D$-dimension. We may either directly carry out Baikov integrals
  with non-maximal cut, or extend known maximal-cut functions to the
  non-maximal cut results via the block-triangular differential equation. 
\end{itemize}

%First of all, it is interesting to examine
%more complicated kinematic configurations giving rise to generalizations of%
%Gauss' hypergeometric functions. For example, the maximal cut of the two-ass
%short-side planar double box is the Appell F1 function. Moreover, the Baikov
%representation is straightforward to extend beyond two loops. We anticipate
%that our procedure can be generalized realtively quick to such cases.

\acknowledgments{We thank S. Badger, N. Beisert, C. Duhr,
  H. Frellesvig, A. Georgoudis, J. Henn,
  H. Ita, D. Kosower, K. Larsen, R. Lee, Mastrolia, E. Panzer, C. Papadopoulos,
  S. Weinzierl and
  M. Zeng for enlightening discussions. Furthermore, we acknowledge
  K. Larsen for participation in the early stage of this work. 
The work of J.B. is supported by the Swiss National Science Foundation through the NCCR SwissMap.
M.S. is a Sapere Aude fellow supported by the Danish Council
for Independent Research under contract No. DFF-4181-00563. Y.Z. is
a Swiss National Science Foundation Ambizione fellow with the grant
No. PZ00P2\_161341. }

\appendix

\section{Rudiments of hypergeometric identities}
In this appendix, we list several identities for hypergeometric
functions which are used in this paper. The discussion follows \cite{MR1424469,MR1034956}.

Hypergeometric $\,_2F_1$ functions are the solutions of Fuchsian
equations with three regular singular points on $\mathbb {CP}^1$. With
Taylor series, it is defined as 
\begin{equation}
  \label{eq:47}
  \,_2F_1(\alpha,\beta,\gamma,z) = \sum_{n=0}^{\infty}
  \frac{(\alpha)_n(\beta)_n}{n! (\gamma)_n} z^n,\quad |z|<1\,,
\end{equation}
where the Pochhammer symbol \eqref{eq:Pochhammer} is used. For other points (except $z=1$) on the complex plane, $\,_2F_1(z)$ can
be defined by analytic continuation. %along paths which avoid $0$
%and $1$. 

$\,_2F_1(z)$ satisfies the Fuchsian equation (when $\gamma\not \in
\mathbb Z_{\leq 0}$),
\begin{equation}
  \label{hyp2F1DE}
  z(1-z)\frac{d^2 f}{d z^2} +\big(\gamma-(\alpha+\beta+1)z\big)
  \frac{d f}{dz} -\alpha\beta f=0\,.
\end{equation}
The other independent solution is (when $\gamma\not \in
\mathbb Z$),
\begin{equation}
  \label{eq:49}
  z^{1-\gamma} \,_2F_1 (\alpha-\gamma+1,\beta-\gamma+1,2-\gamma,z)\,.
\end{equation}

It is possible to study the solution of \eqref{hyp2F1DE} near the
other singular two points $1$ and $\infty$, and the solution will be
hypergeometric functions with the fourth arguments $1-z$ and $1/z$. By
the linear dependence of solutions, there are relations,
\begin{align}
  \,_2F_1(\alpha,\beta,\gamma,z) &=\frac{\Gamma (\gamma ) \Gamma (\alpha +\beta -\gamma )
    (1-z)^{-\alpha -\beta +\gamma } \, _2F_1(\gamma -\alpha ,\gamma
    -\beta ;-\alpha -\beta +\gamma +1;1-z)}{\Gamma (\alpha ) \Gamma
    (\beta )}\nonumber \\
&+\frac{\Gamma (\gamma ) \Gamma (-\alpha -\beta +\gamma ) \,
  _2F_1(\alpha ,\beta ;\alpha +\beta -\gamma +1;1-z)}{\Gamma (\gamma
  -\alpha ) \Gamma (\gamma -\beta )}\,, \label{HypShift}\\
\,_2F_1(\alpha,\beta,\gamma,z) &=\frac{\Gamma (\gamma ) (-z)^{-\alpha }
  \Gamma (\beta -\alpha ) \, _2F_1\left(\alpha ,\alpha -\gamma
    +1;\alpha -\beta +1;\frac{1}{z}\right)}{\Gamma (\beta ) \Gamma
  (\gamma -\alpha )}\nonumber \\
&+\frac{\Gamma (\gamma ) (-z)^{-\beta } \Gamma (\alpha -\beta ) \,
  _2F_1\left(\beta ,\beta -\gamma +1;-\alpha +\beta
  +1;\frac{1}{z}\right)}{\Gamma (\alpha ) \Gamma (\gamma -\beta )}\,.
\label{HypInv}
\end{align}
Euler's transformation of hypergeometric $\,_2F_1$ functions is
\begin{gather}
  \,_2F_1(\alpha,\beta_,\gamma_,z) =(1-z)^{\gamma-\alpha-\beta}
  \,_2F_1(\gamma-\alpha, \gamma-\beta_,\gamma_,z)\,.
\label{HypEuler}
\end{gather}
When calculating the maximal cut functions of different subregions, we frequently use
\eqref{HypShift}, \eqref{HypInv} and \eqref{HypEuler} to connect
various $\,_2F_1$ functions.

The functions $\,_2F_1(\alpha+l,\beta+m,\gamma+n,z)$, with $l,m,n\in
\mathbb Z$ are called ``contiguous'' to the function
$\,_2F_1(\alpha,\beta,\gamma,z)$. Any three contiguous $\,_2F_1$ functions
satisfy Gauss' contiguous relations, 
\begin{gather}
  \label{eq:50}
  A_1 F_1+A_2 F_2+A_3 F_3=0\,,
\end{gather}
where the coefficients $A_i$'s are rational functions in $\alpha$, $\beta$, $\gamma$ and
$z$. The two fundamental Gauss' contiguous relations follow from the
integral representation of $\,_2F_1(z)$,
\begin{gather}
  \label{eq:51}
  (\gamma -1) \,_2F_1(\alpha,\beta,\gamma-1,z) -\alpha
  \,_2F_1(\alpha+1,\beta_,\gamma_,z)-(\gamma-\alpha-1)\,_2F_1(\alpha,\beta_,\gamma_,z)=0\,\\
\gamma  \,_2F_1(\alpha,\beta,\gamma,z) -\beta z
\,_2F_1(\alpha,\beta+1,\gamma+1,z)-\gamma \,_2F_1(\alpha-1,\beta,\gamma,z)=0\,,
\end{gather}
and all other Gauss' contiguous relations can be derived from these
two. We use these relations for studying the IBPs and dimension-shift identities on the maximal cut
level. 

When studying the maximal cut in arbitrary dimension, we also meet
generalized hypergeometric functions, for example, the Appell F1
function. It has four parameters and two variables, and is defined as, 
\begin{equation}
  \label{eq:61}
  F_1(\alpha;\beta_1,\beta_2;\gamma;x,y)=\sum_{m=0}^\infty
  \sum_{n=0}^\infty \frac{(\alpha)_{m+n}(\beta_1)_m (\beta_2)_n}{m!n!
    (\gamma)_{m+n}} x^m y^n,\quad |x|<1,|y|<1.
\end{equation}
It can be defined outside the polydisc by analytic continuation.

One-variable integrals with four distinct factors can be expressed as the Appell F1
function, for instance,
\begin{gather}
  \label{eq:62}
\int_0^1 u^{\alpha-1} (1-u)^{\gamma-\alpha-1} (1-ux)^{-\beta_1}
(1-uy)^{-\beta_2} du
=  \frac{\Gamma(\alpha)\Gamma(\gamma-\alpha)}{\Gamma(\gamma)} F_1(\alpha;\beta_1,\beta_2;\gamma;x,y)\,.
\end{gather}
The contiguous relations for Appell F1 functions can be found, for example, in the survey
article \cite{Schlosser2013}.

\bibliographystyle{elsarticle-num}
\bibliography{D_cut}
\end{document}